\documentclass{aa}  

\usepackage{graphicx}
\usepackage[colorlinks, linkcolor=blue, anchorcolor=blue, citecolor=blue]{hyperref}
\usepackage{txfonts}
\usepackage{color}
\usepackage{siunitx}
\usepackage{mathtools}
\usepackage{amsmath}
\usepackage{CJK}

\usepackage{natbib}
\bibpunct{(}{)}{;}{a}{}{,} % to follow the A&A style

\begin{document}

   \title{The crater-induced YORP effect}

%   \subtitle{I. Overviewing the $\kappa$-mechanism}

   \author{Wen-Han Zhou 
          \inst{1,2}
          \and
          Yun Zhang
          \inst{3}
          \and
        Xiaoran Yan 
          \inst{1,4}
          \and
          Patrick Michel
          \inst{1}
          }

   \institute{Universit\'e C\^ote d'Azur, Observatoire de la C\^ote d'Azur, CNRS, Laboratoire Lagrange, Nice 06304, France\\
              \email{wenhan.zhou@oca.eu}
        \and
        Origin Space Co. Ltd., China
        \and
             Department of Aerospace Engineering, University of Maryland, College Park, MD, 20742, USA
        \and
             Department of Aerospace Engineering, Tsinghua University, Beijing, China
             }

% \abstract{}{}{}{}{} 
% 5 {} token are mandatory
 
  \abstract
   {The Yarkovsky$-$O'Keefe$-$Radzievskii$-$Paddack (YORP) effect plays an important role in the rotational properties and evolution of asteroids. While the YORP effect induced by the macroscopic shape of the asteroid and by the presence of surface boulders has been well studied, no investigation has been performed yet regarding how craters with given properties influence this effect.}
   {We introduce and estimate the crater-induced YORP effect (CYORP), which arises from the concave structure of the crater, to investigate the magnitude of the resulting torques as a function of varying properties of the crater and the asteroid by a semi-analytical method.}
   {By using a simple spherical shape model of the crater and assuming zero thermal inertia, we calculated the total YORP torque due to the crater, which was averaged over the spin and orbital motions of the asteroid, accounting for self-sheltering and self-sheltering effects.}
   {The general form of the CYORP torque can be expressed in terms of the crater radius $R_0$ and the asteroid radius $ R_{\rm ast}$: $<T_{\rm CYORP}> \sim W  R_0^2  R_{\rm ast} \Phi/c,$ where $W$ is an efficiency factor. We find that the typical values of $W$ are about 0.04 and 0.025 for the spin and obliquity component, respectively, which indicates that the CYORP can be comparable to the normal YORP torque when the size of the crater is about {one-tenth} of the size of the asteroid, or equivalently when the crater/roughness covers {one-tenth} of the asteroid surface. Although the torque decreases with the crater size $R_0$ as $\sim R_0^2$, the combined contribution of all small craters can become non-negligible due to their large number when the commonly used power-law crater size distribution is considered. The CYORP torque of small concave structures, usually considered as surface roughness, is essential to the accurate calculation of the complete YORP torque. Under the CYORP effect that is produced by collisions, asteroids go through a random walk in spin rate and obliquity, with a YORP reset timescale typically of 0.4~Myr. This has strong implications for the rotational evolution and orbital evolution of asteroids.}
   % Conclusion
   {Craters and roughness on asteroid surfaces, which correspond to concave structures, can influence the YORP torques and therefore the rotational properties and evolution of asteroids. We suggest that the CYORP effect should be considered in the future investigation of the YORP effect on asteroids.}

   \keywords{minor planets, asteroids: general}

   \maketitle
%
%-------------------------------------------------------------------

\section{Introduction}

 The Yarkovsky-O'Keefe-Radzievskii-Paddack (YORP) effect, which is a thermal torque produced by surface emission, has a strong influence on the rotational state and evolution of asteroids \citep{Rubincam00, Vokrouhlicky02,Bottke06}. It can either increase or decrease the spin rate and can also change the spin obliquity of an asteroid on timescales that also depend on physical and dynamical properties of the considered asteroid \citep[e.g.,][]{Capek2004, Scheeres2008, Statler2009, Rozitis2012}. Although a slow process in general, it could be directly measured by ground-based observations \citep[e.g.,][]{Vdurech18}. Moreover, it provides an explanation for some observed properties, such as the preferred orientation of the spin axis of members of the Koronis asteroid family \citep{Vokrou2003}, as well as some asteroid shapes, such as the top shapes of primaries of small binary systems \citep[e.g.,][]{Walsh2008}, and possibly the shapes of  the asteroids Bennu and Ryugu \citep[although another explanation has been proposed for these particular cases;][]{Michel2020}. 
 
 In particular, to be at the origin of top-shaped asteroids, the YORP effect needs to cause an increase in spin rate on a continuous basis or in a trend that allows the shape to evolve in a spinning top on a timescale that makes it possible. However, it was found that small changes in the surface topography of an asteroid can strongly influence the YORP effect outcome \citep{Statler2009}, for instance, causing a spin down rather than a spin up, which could alter a systematic increase in the rotation rate and potentially make it difficult to achieve a top shape. Therefore, it is crucial to assess the effect of surface topography on the total YORP torque. 
 
 The current YORP model reads
\begin{equation}
\label{eq:T_YORP1}
    \vec T_{\rm YORP} = \vec T_{\rm NYORP} + \vec T_{\rm TYORP},
\end{equation}
where $\vec T_{\rm NYORP}$ stands for the YORP effect on the whole asteroid, and $\vec T_{\rm TYORP}$ stands for the tangential YORP effect, which describes the YORP effect related to the presence of boulders and surface roughness \citep{Golubov12,Golubov14, Golubov17}. 

\begin{figure*}
    \centering
    \includegraphics[width = \textwidth]{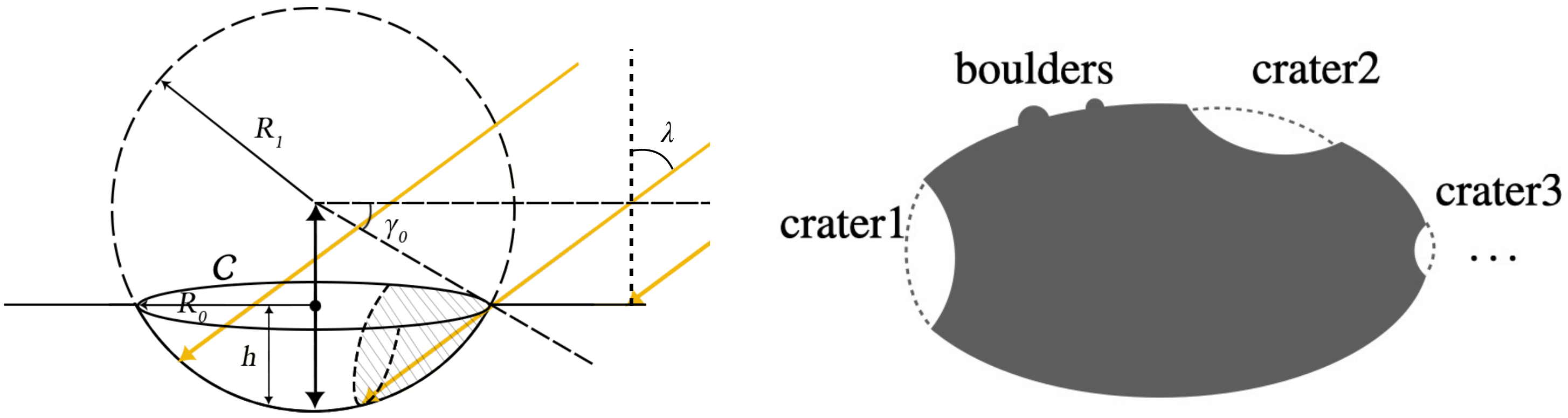}
    \caption{Simple crater model (left panel) and features in the asteroid that affect the total YORP torque (right panel). In the left panel, the yellow arrow represents the light coming from the direction of the Sun, and the shadow region is shaded in gray. Parameters $h$ and $R_0$ are the depth and radius of the crater, respectively. The edge circle of the crater on the ground level is denoted by $\mathcal{C}$. The angle between the sunlight and ground normal is denoted by $\lambda_0$. The right panel shows the boulders and craters on the surface of asteroids that could contribute to the total YORP effect.}
    \label{fig1}
\end{figure*}
 
 Here, we consider another surface characteristic that has not been considered so far and that might also influence the evolution of an asteroid rotation state under the YORP effect. Images sent by space missions showed us that asteroid surfaces are populated with craters, whose distribution and properties can differ from one object to the next \citep[see, e.g.,][for a review]{Marchi2015}, depending on its age, its response to impacts, and other possible processes, such as surface motions that can erase small features or boulder armoring that can prevent a crater from forming \citep[e.g.,][]{Bierhaus2022,Daly2022}. Nevertheless, craters are an important and systematic characteristic of asteroid surfaces that may have some influence on the YORP effect because this effect is sensitive to the fine topography \citep{Statler2009}. 
 
For the first time, we propose here the concept of the crater-induced YORP (called CYORP hereafter) and show that CYORP may contribute to the total YORP torque as well, which adds a {"CYORP"} term into Equation~(\ref{eq:T_YORP1}), 
\begin{equation}
    \vec T_{\rm YORP} = \vec T_{\rm NYORP} + \vec T_{\rm TYORP} + \vec T_{\rm CYORP},
\end{equation}
where 
\begin{equation}
\label{eq:T_CYORP_total}
    \vec T_{\rm CYORP,total} = \Sigma_i T_{{\rm CYORP},i}
\end{equation}
as a summation for a whole set of craters or concave structures on the asteroid (see Fig.~\ref{fig1}). The CYORP torque is the difference between the torque caused by the crater and the torque by the ground before the birth of the crater,
\begin{equation}
\label{eq:T_CYORP1}
    \vec T_{\rm CYORP} = \vec T_{\rm crater} - \vec T_{\rm ground}.
\end{equation}
Here $\vec T_{\rm ground}$ is the normal YORP torque of the ground before the birth of the crater (see Fig.~\ref{fig1}), which can be expressed as
\begin{equation}
\label{eq:T_ground}
    \vec T_{\rm ground} = \pi R_0^2 \frac{2\Phi}{3c} \cos \lambda \vec r_0 \times \vec n_0,
\end{equation}
where $R_0$ is the radius of the crater, $\Phi$ is the solar flux on the asteroid, $c$ is the speed of light, $\lambda$ is the incident angle of the light, and $\vec r_0$ and $\vec n_0$ are the position vector and unit normal vector of the crater, respectively (see Fig.~\ref{fig2}).

The CYORP torque arises due to the concave structure of the crater. The vertical wall of the crater induces a force tangential to the surface, and the curvature of the crater induces a normal force component that is different from the force that is produced by the ground without the crater. Thus, the force that leads to the CYORP torque comprises of both the tangential and normal components. The self-sheltering and self-heating effects because of concavity influence the total torque; this is also considered in this work. 
% The self-sheltering and self-heating are the main causes of the net torque. Therefore, the CYORP effect has to be distinguished from the TYORP effect, although both describe the fine structure of the surface. For example, one difference is the CYORP torque still exists in the zero-thermal-inertia regime while the TYORP does not. 
In general, $\vec T_{\rm CYORP}$ takes the form of the following scaling rule with the radius of the crater $R_0$ and of the asteroid $ R_{\rm ast}$:
\begin{equation}
    \vec T_{\rm CYORP} = W\frac{\Phi}{c}R_0^2  R_{\rm ast},
\end{equation}
where $W$ is a function of the properties of the crater and the asteroid (the detailed derivation of this equation is presented in Section~\ref{sec2}). As a general rule, $\vec T_{\rm CYORP}$ is thus proportional to the square of the crater radius and to the asteroid radius. Based on this scaling relation, we developed a semi-analytic method that can be applied to the calculation of the CYORP effect, and it provides a basic understanding of the relative influence of each parameter. The derived CYORP torque can be applied both for craters and for any concave structures on the surface of an asteroid, although a modification accounting for the geometry is needed.

 We focus here on one crater and vary its properties to determine how they influence the YORP effect. As a first step, we assume zero thermal inertia \citep[Rubincam's approximation; see][]{Rubincam00}, which can be applied to asteroids with low thermal conductivity or slow rotation. Rubincam's approximation is suitable for calculating the spin component of the YORP torque. The model including the thermal inertia of the asteroid will be the topic of a next study. In the following, we present our calculation of the crater-induced YORP torque in Section~\ref{sec2}, accounting for the crater shape and other related thermophysical processes. Section~\ref{sec3} presents results for various asteroid properties and locations of the crater. In Section~\ref{sec4} we give the typical value of the CYORP torque (Sec. \ref{sec4_1}), which could be used to estimate the order of magnitude, and we analyze the applicability of the CYORP effect to the complete YORP torque and to the spin evolution of asteroids (Sec. \ref{sec4_2}). In Section~\ref{sec5} we summarize the main results and draw the conclusion.
\section{Calculation of the crater-induced YORP torque}
\label{sec2}

% \subsection{heat model}
% The radiation-induced recoil force consists of three parts, which are caused by the incident, scattered, re-emitted photons, respectively. It has been proven that the first part does not contribute to the YORP torque during the orbital motion. Hence, we consider only the recoil forces caused by scattered and re-emitted photons.
% {\color{blue} more details later on}

\subsection{Shape model for the crater}
We considered a simple shape model for the crater, which is represented by a full or part of a semi-sphere with a radius $R_1$ and depth $h$ (see Fig.~\ref{fig1}). In this way, the size and the shape of the crater can be determined by two parameters $R_1$ and $\gamma_0$, where
\begin{equation}
    \sin \gamma_0 = \frac{R_1 - h}{R_1}.
\end{equation}
We considered a coordinate system ($x$, $y$, $z$) with the origin located at the sphere center (see Fig.~\ref{fig2}). The unit vectors $\vec e_x$, $\vec e_y $ , and $\vec e_z$ were chosen so that $\vec e_z$ lay along the symmetry axis of the spherical crater and $\vec e_x$ lay in the plane of $\vec e_z$ and the unit solar position vector $\vec s$. Vector $\vec e_y$ follows the right-hand rule. Equivalently, $\vec e_y$ and $\vec e_z$ are defined as
\begin{equation}
    \begin{aligned}
        & \vec e_y = \vec e_z \times \vec s, \\
        & \vec e_x = \vec e_y \times \vec e_z.
    \end{aligned}
\end{equation}
In this coordinate system (see Fig.~\ref{fig2}), the crater can be defined as
\begin{equation}
\label{eq:crater}
    \mathcal{Z} \coloneqq \{ (x,y,z)  \in \mathbb{R}^3 \lvert  x^2+y^2+z^2 =R_1, z \geq R_1 \sin \gamma_0 \}.
\end{equation}
Applying
\begin{equation}
\label{eq:xyz}
    \left \{
    \begin{aligned}
    & x = r\sin \theta \cos \phi \\
    & y = r\sin \theta \sin \phi \\
    & x = r\cos \theta,
    \end{aligned}
    \right.
\end{equation}
the crater is equivalently
\begin{equation}
    \mathcal{Z} \coloneqq \{ (x,y,z)  \in \mathbb{R}^3 \lvert  r =R_1, \theta \in (0,\pi/2 - \gamma_0), \phi \in (0,2\pi) \}.
\end{equation}
The widely used parameter depth-diameter ratio translates as
\begin{equation}
    \frac{h}{D_0} = \frac{1 - \sin \gamma_0}{2 \cos \gamma_0},
\end{equation}
where $D_0 = 2R_0$ is the diameter of the crater.
% {\color{blue} There needs some information of real asteroids}.

\subsection{Shadowing effect}
For a concave geometry such as a crater, the influence of self-shadowing plays a significant role for the YORP effect. There are three consequences of self-shadowing. (1) The crater is sheltered by itself, that is, the fraction in the shadow of the crater does not receive the photons from direct solar radiation. (2) The effective angular momentum transfer that occurs in a surface element is affected by the neighboring topology because the radiated photons can be reabsorbed by the shelter. As a result, the effective recoil force (and the YORP torque) is different from that in the case of a nonsheltered environment. (3) The radiation caused by secondary illumination from the crater itself, which is ignored in this work for simplicity. As we show in Section~\ref{sec3}, the first effect of shadowing contributes to the net YORP torque of a crater, and the second effect weakens the YORP torque.

\subsubsection{Illuminated area}
\label{sec2_2_1}
The unit vector directed toward the Sun from the crater $\vec s'$ represents the direction of the parallel sunlight. When we consider that the size of a typical asteroid $ R_{\rm ast}$ (on the order of some $\text{kilometers}$) is much smaller than the distance from the Sun $d$ (on the order of one au), the unit vector pointing from the center of the asteroid to the Sun is $\vec s \simeq \vec s'$. In the following context, we use $\vec s$ to denote the position of the Sun relative to both the asteroid and the crater.

The unit position vector of the Sun in the coordinate system ($x$, $y$, $z$) can be expressed as
\begin{equation}
\label{eq:s}
    \vec s = \sin \lambda \vec e_x - \cos \lambda \vec e_z,
\end{equation}
where $\lambda$ is the incident angle of the light.
% The unit normal vector of the surface element where $P$ is located is
% \begin{equation}
%     \vec n_p = - \vec r_p.
% \end{equation}
To determine the region that is exposed to sunlight, we need to find the expression function of the boundary of the illuminated region. First, we define the edge of the crater at the ground level, which is a circle as
\begin{equation}
    \mathcal{C} \coloneqq \{ (x,y,z) \lvert x^2+y^2 = R_1^2 \cos^2\gamma_0, z = R_1 \sin \gamma_0 \}.
\end{equation}

\begin{figure*}
    \centering
    \includegraphics[width = \textwidth]{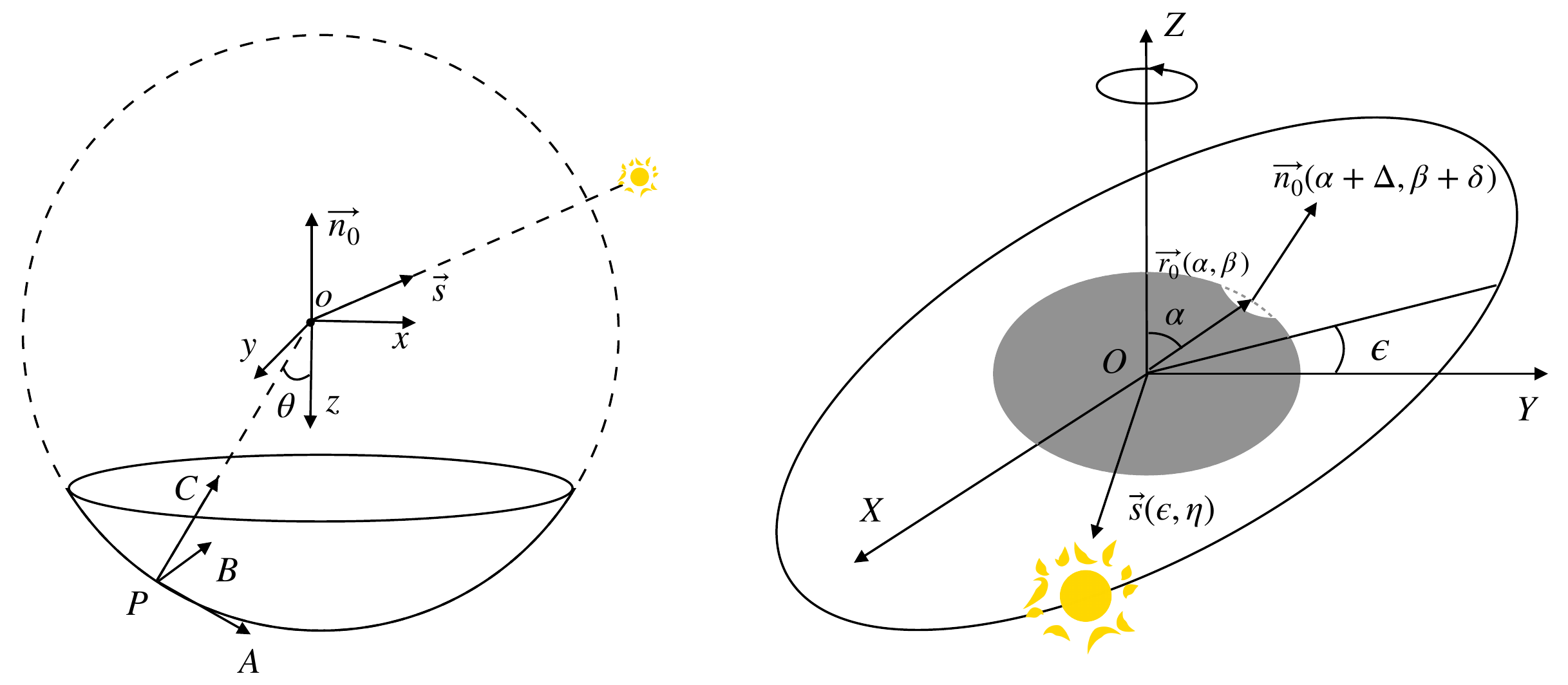}
    \caption{Three coordinate systems in this paper: Coordinate system $oxyz$ for calculating the illuminated domain in the crater, $PABC$ for calculating the effective recoil force of an arbitrary surface element, and $OXYZ$ for averaging the YORP torque over the spin and orbital motion.}
    
    % The coordinate system $PABC$ (left) and $OXYZ$ (right). The axis $Pc$ is the normal vector of point $P$, $Pa$ is at the plane of $Pc$ and $\vec z$ and $Pb$ follows the right hand rule. The center of $OXYZ$ is at the mass center of the asteroid. The $OZ$ axis denoting the major principal axis of the asteroid. The relative position of the Sun is shown with $\epsilon$ denoting the obliquity of the asteroid. The angle $\beta$ and $\eta$ are the spin and orbital angles, respectively.}
    \label{fig2}
\end{figure*}

The boundary is the projection of the upmost circle $C_1$ of the crater on the crater $\mathcal{Z}$ along the light. The boundary can be obtained by solving the intersection of the crater $\mathcal{Z}$ and an elliptic cylinder, which contains $C_1$ and along $\vec s$. When an arbitrary point in the circle $C_1$ is $(x',y',z')$, the elliptic cylinder is
\begin{equation}
    \left \{
    \begin{aligned}
    & x'^2 + y'^2 = R_1^2\cos^2 \gamma_0 \\
    & z' = 0 \\
    & \frac{x-x'}{\sin \lambda} = \frac{z-z'}{-\cos \lambda} \\
    & y = y'.
    \end{aligned}
    \right.
\end{equation}
After reduction, the expression of this elliptic cylinder is
\begin{equation}
    (x+ z\tan \lambda)^2 +y^2 = R_1^2.
\end{equation}
Combining this with the expression of the crater (Equation~(\ref{eq:crater})) and applying $x = R_1\sin \theta \sin \phi$, $y =R_1\sin \theta \cos \phi $ and $z = R_1\cos \theta$, we obtain the expression of the intersection curve,
\begin{equation}
\label{eq:cos_phi}
    \cos \phi = \frac{\cos 2 \lambda \cos \theta+ \sin \gamma_0}{\sin 2\lambda \sin \theta}.
\end{equation}
Because $0 < \phi < 2 \pi$, given a polar angle $\theta$, Equation~(\ref{eq:cos_phi}) has two solutions (if a solution exists) $\phi_1$ and $\phi_2$ , for $\phi$ with $\phi_1 + \phi_2 = 2\pi$, in which we assume $\phi_1 < \phi_2$ for further analysis. The illuminated region  is represented by
\begin{equation}
\label{eq:W}
    \mathcal{W} \coloneqq \left\{ (x,y,z) \in \mathcal{Z} \lvert  \cos \phi < \frac{\cos 2 \lambda \cos \theta+ \sin \gamma_0}{\sin 2\lambda \sin \theta} \right\}.
\end{equation}
% or equivalently, 
% \begin{equation}
%     \mathcal{W} \coloneqq \left\{ (x,y,z) \in \mathcal{Z} \lvert  \phi_1 < \phi < \phi_2 \right\}.
% \end{equation}
% which will be used for integration of the torque in Chapter \ref{sec2_4}.

It is not guaranteed that Equation~(\ref{eq:cos_phi}) has a solution because the right side of the equation can be larger than 1. Depending on the incident angle of the light $\lambda$, there are three illumination modes, given an incident angle of light $\lambda$ as follows:
\begin{description}
\item[(1)] The whole crater is illuminated $ \iff \lambda < \gamma_0$. In this case,
$\mathcal{W} = \mathcal{Z.}$
\item[(2)] Two sides of the crater (e.g., east and west) are illuminated $\iff \gamma_0 < \lambda < \pi/4 + \gamma_0/2$. In this case, $\mathcal{W} = \{ (x,y,z) \in \mathcal{Z} \lvert  \theta \in (0,\pi/2-2\lambda+\gamma_0), \phi \in (0,2 \pi) \, \, {\rm or} \,\,\theta \in (\pi/2-2\lambda+\gamma_0 , \pi/2 - \gamma_0), \phi \in (\phi_1 ,\phi_2) \}.$
\item[(3)] One side of the crater is illuminated $\iff \pi/4 + \gamma_0/2 < \lambda < \pi/2 $. In this case, $\mathcal{W} = \{ (x,y,z) \in \mathcal{Z} \lvert  \theta \in (2\lambda - \gamma_0 - \pi/2, \pi/2 - \gamma_0), \phi \in (\phi_1 ,\phi_2)\}.$
\end{description}
We refer to Appendix \ref{appA} for the details of the above mathematical description of these three illumination modes, or a self-examination may be made through plane geometry in Fig \ref{fig1}. 
Although the different illumination modes are based on Equation~(\ref{eq:W}), they refer to different integration domains. Expressing them explicitly helps solve the thermal recoil force of the crater (see Eq.~(\ref{eq:F}) in  Section~\ref{sec2_2_2}), as the crater could go through all these modes during a rotational period.

\subsubsection{Self-heating}

\label{sec2_2_2}
The surface of asteroids experiences three types of forces, which are caused by absorbed, scattered, and reemitted photons, respectively. The torque produced by absorbed photons is proven to average out after integrating over the spin and orbital periods for any asteroid shapes. Therefore, this type of force does not contribute to the YORP torque of the whole asteroid \citep{Nesvorny2008}. Both the recoil forces produced by scattered and reemitted photons depend on the light scattering law. We assumed the simple and most widely used Lambert scattering law, in which the light is emitted in all directions with an intensity proportional to the cosine of the angle between the light direction and the normal vector of the surface. For concave configurations such as a crater, we should consider the self-heating effect, which arises because some of the photons may be re-absorbed by the nearby shelter, which prevents them from contributing to the effective recoil force \citep{Statler2009, Yan2019}. The reemission of the obscuring parts, which might make a difference for a concave structure \citep{Rozitis2013}, is ignored in this work for its complexity and will be studied in the future.

We consider a reference frame with the origin at the point $P$ located at the polar angle $\theta$ (see Fig. \ref{fig2}). The orthogonal basis is represented by three unit vectors:
\begin{equation}
\label{eq:abc}
    \begin{aligned}
    & \vec e_C = \vec n \\
    & \vec e_B = \vec e_C \times \vec e_z \\
    & \vec e_A = \vec e_B \times \vec e_C.
    \end{aligned}
\end{equation}
Here $\vec n$ is the normal vector of the surface element. An arbitrary vector of the light ray that is emitted through a solid angle d$\Omega$ can be expressed by the polar angle $\mu$ and the azimuth angle $\nu$ in this reference frame. According to Lambert's scattering law, the recoil force on the surface element d$S$ that is located at the latitude of $\theta$ can be expressed as 
\begin{equation}
    \vec{f} = -\int_{\mathcal{H}} \frac{\Phi}{c} \cos \mu
    \left (
    \begin{aligned}
    &\sin \mu \cos \nu \\
    &\sin \mu \sin \nu \\
    &\cos \mu
    \end{aligned}
    \right ) {\rm d}\Omega.
\end{equation}
Here $\mathcal{H}$ is the region on the sky in which the light ray is not reabsorbed. In the spherical coordinate system, the boundary of $\mathcal{H}$ is the intersection curve of an elliptic cone and the unit sphere, both of which are centered at the origin. This elliptic cone must contain the upmost circle ($\mathcal{C}$ in Figure~\ref{fig1}). When an arbitrary point in $\mathcal{C}$ is ($x',y',z'$), the elliptic cone expressed by ($x,y,z$) is
\begin{equation}
    \left \{
    \begin{aligned}
    & y'^2 +  [(x'- x_0)\cos \theta - (z'- z_0)\sin \theta]^2  = \cos^2 \gamma_0 \\
    & (x'- x_0)\sin \theta +(z'-z_0)\cos \theta = 0 \\
    & \frac{x}{x'} = \frac{y}{y'} =\frac{y}{y'} 
    \end{aligned}
    \right.
.\end{equation}
Here $(x_0,0,z_0)=(\sin \gamma_0 \sin \theta, 0, 1- \sin \gamma_0 \cos \theta_0) $ is the center of the circle $\mathcal{C}$ in the coordinate system ($a,b,c$). Combining this with the unit sphere, which is
\begin{equation}
    \left \{
    \begin{aligned}
    & x  = \sin \mu \cos \nu \\
    & y  = \sin \mu \sin \nu \\
    & z  = \cos \mu ,
    \end{aligned}
    \right.
\end{equation}
we obtain the function of the boundary of the illuminated sky:
\begin{equation}
    \cos \nu = -\frac{\cos 2 \mu \cos \theta + \sin \gamma_0}{\sin 2\mu \sin \theta}.
\end{equation}
Here we assume that the solutions of this equation for $\nu$ are $\nu_1$ and $\nu_2$ with $\nu_1<\nu_2$. Thus, the region $\mathcal{H}$ is 
\begin{equation}
\label{eq:H}
    \mathcal{H} = \left\{ (x,y,z) \in \mathbb{R}^3 \lvert  \mu \in (0,\pi/2),\cos \nu < -\frac{\cos 2 \mu \cos \theta+ \sin \gamma_0}{\sin 2\mu \sin \theta,} \right\}.
\end{equation}
It can be also expressed as
\begin{equation}
    \mathcal{H} = \left\{ (x,y,z) \in \mathbb{R}^3 \lvert  \nu \in (0,2\pi),\mu \in (\frac{\pi + \gamma_0 - \gamma_1}{2} ,\pi) \right\},
\end{equation}
where
\begin{equation}
    \gamma_1 = \arctan \left( \frac{1}{\tan \theta \cos \nu} \right) \in (0, \pi).
\end{equation}

% \begin{equation}
%     \mathcal{H} = \mathcal{H}_1 \cup \mathcal{H}_2,
% \end{equation}
% where
% \begin{equation}
%     \begin{aligned}
%     & \mathcal{H}_1 = \left\{  (x,y,z) \in \mathbb{R}^3 \lvert  \mu-\frac{\pi}{4} \in (\frac{\gamma_0+\theta}{2},\frac{\pi}{4}), \nu \in (0, 2\pi)    \right\} \\
%     & \mathcal{H}_2 = \left\{  (x,y,z) \in \mathbb{R}^3 \lvert  \mu-\frac{\pi}{4} \in (\frac{\gamma_0-\theta}{2},\frac{\gamma_0+\theta}{2}), \nu \in (\nu_1, \nu_2)    \right\} \\.
%     \end{aligned}
% \end{equation}
Therefore, the recoil force of the surface element that is located at the latitude $\theta$ is
\begin{equation}
\begin{aligned}
    \vec{f} = -\int_{\mathcal{H}} \frac{\Phi}{c} \cos \mu
    \left (
    \begin{aligned}
    &\sin \mu \cos \nu \\
    &\sin \mu \sin \nu \\
    &\cos \mu
    \end{aligned}
    \right ) \sin \mu {\rm d}\nu {\rm d}\mu = 
    \left (
    \begin{aligned}
    & f_1'(\theta) \\
    & 0 \\
    & f_3'(\theta)
    \end{aligned}   
    \right ).
\end{aligned}
\end{equation}
Here $f_1'(\theta)$ and $f_3'(\theta)$ are functions of the latitude $\theta$ of the surface element, resulting from $\nu_1$ and $\nu_2$. The second component cancels out because the integral of $\sin \nu$ over either ($\nu_1,\nu_2$) or ($0,2\pi$) is zero. Given $\vec e_C = \vec n$ and $\vec e_A = \vec n /\tan \theta + \vec e_z/\sin \theta $ (see Eq.~(\ref{eq:abc})), we have
\begin{equation}
    \vec f = f_1(\theta) \vec n + f_3(\theta) \vec e_z ,
\end{equation}
with
\begin{equation}
\begin{aligned}
    & f_1(\theta) =  f_3'(\theta)+\frac{f_1'(\theta)}{\tan \theta}, \\
    & f_3(\theta) =  \frac{f_3'(\theta)}{\sin \theta}.
\end{aligned}
\end{equation}

\subsection{Integral of the recoil force}

The total recoil force of the crater can be obtained by integrating the recoil force (Sec.~\ref{sec2_2_2}) over the illuminated region (Sec.~\ref{sec2_2_1}),
\begin{equation}
    \vec F = \int_{\mathcal{W}} \vec f {\rm d}S = \int_{\mathcal{W}} f_1(\theta) \vec n + f_3(\theta) \vec e_z \sin \theta {\rm d} \phi {\rm d}\theta.
\end{equation}
Because $\vec n = -\sin \theta \cos \phi \vec e_x - \sin \theta \sin \phi \vec e_y - \cos \theta \vec e_z$, in the coordinate system $(x,y,z)$, we have
\begin{equation}
\label{eq:F}
    \vec{F} = \int_{\mathcal{W}} \sin \theta
    \left (
    \begin{aligned}
    & -f_1(\theta) \sin \theta \cos \phi \\
    & -f_1(\theta) \sin \theta \sin \phi\\
    & -f_1(\theta) \cos \theta + f_3(\theta)
    \end{aligned}
    \right ) {\rm d}\phi {\rm d}\theta = 
    \left(
    \begin{aligned}
    & F_1(\lambda) \\
    & 0 \\
    & F_3(\lambda) 
    \end{aligned}
    \right)
.\end{equation}
The $y$-component vanishes due to the symmetry of the integral domain on $\phi$ (see Eq.~\ref{eq:W}). The tangential component and the normal component of the recoil force both exist, and not only one of them as in TYORP or NYORP.

% \begin{equation}
%     \vec F =  \frac{F_1(\lambda)}{\sin \lambda}\vec s - \left( \frac{F_1(\lambda)}{\tan \lambda} + F_3(\lambda) \right) \vec n_0.
% \end{equation}
% This formula will be used in the next section to calculate the torque.

\subsection{Averaged YORP torque}
\label{sec2_4}

The radiative torque is expressed as
\begin{equation}
\label{eq:T_crater1}
    \vec T_{\rm crater} = \int_{\mathcal{W}} \vec r' \times \vec f {\rm d}S .
\end{equation}
Here $\vec r' = \vec r_0 + \vec r$ is the position vector from the mass center of the asteroid to the surface element d$S$ on the crater, where $\vec r_0$ denotes the position vector of the sphere center of the crater, and $\vec r$ denotes the vector from the sphere center to the surface element. Because $\vec r \times \vec f \sim \vec n \times \vec n = 0 $ and $r \ll r_0$, Equation~(\ref{eq:T_crater1}) can be simplified as 
\begin{equation}
\label{eq:T_crater2}
    \vec T_{\rm crater} = \int_{\mathcal{W}} \vec r_0 \times \vec f {\rm d}S = \vec r_0 \times \vec F = F_1(\lambda) \vec r_0 \times  \vec e_x + F_3(\lambda) \vec r_0 \times \vec e_z.
\end{equation}
When we substitute $\vec e_x = \vec s/\sin \lambda + \vec e_z/\tan \lambda $ into Equation~(\ref{eq:T_crater2}), the total torque becomes
\begin{equation}
\label{eq:T_crater3}
    \vec T_{\rm crater} = \frac{F_1(\lambda)}{\sin \lambda} \vec r_0 \times \vec s - \left(\frac{F_1(\lambda)}{\tan \lambda} + F_3(\lambda) \right) \vec r_0 \times \vec n_0.
\end{equation}
Plugging Equations (\ref{eq:T_ground}) and (\ref{eq:T_crater3}) into Equation~(\ref{eq:T_CYORP1}), we obtain 
\begin{equation}
\label{eq:T_CYORP2}
    \vec T_{\rm CYORP} = \frac{F_1(\lambda)}{\sin \lambda} \vec r_0 \times \vec s - \left(\frac{F_1(\lambda)}{\tan \lambda} + F_3(\lambda) + \pi R_0^2 \frac{2\Phi}{3c} \cos \lambda \right) \vec r_0 \times \vec n_0.
\end{equation}

\begin{figure*}
    \centering
    \includegraphics[width=\textwidth]{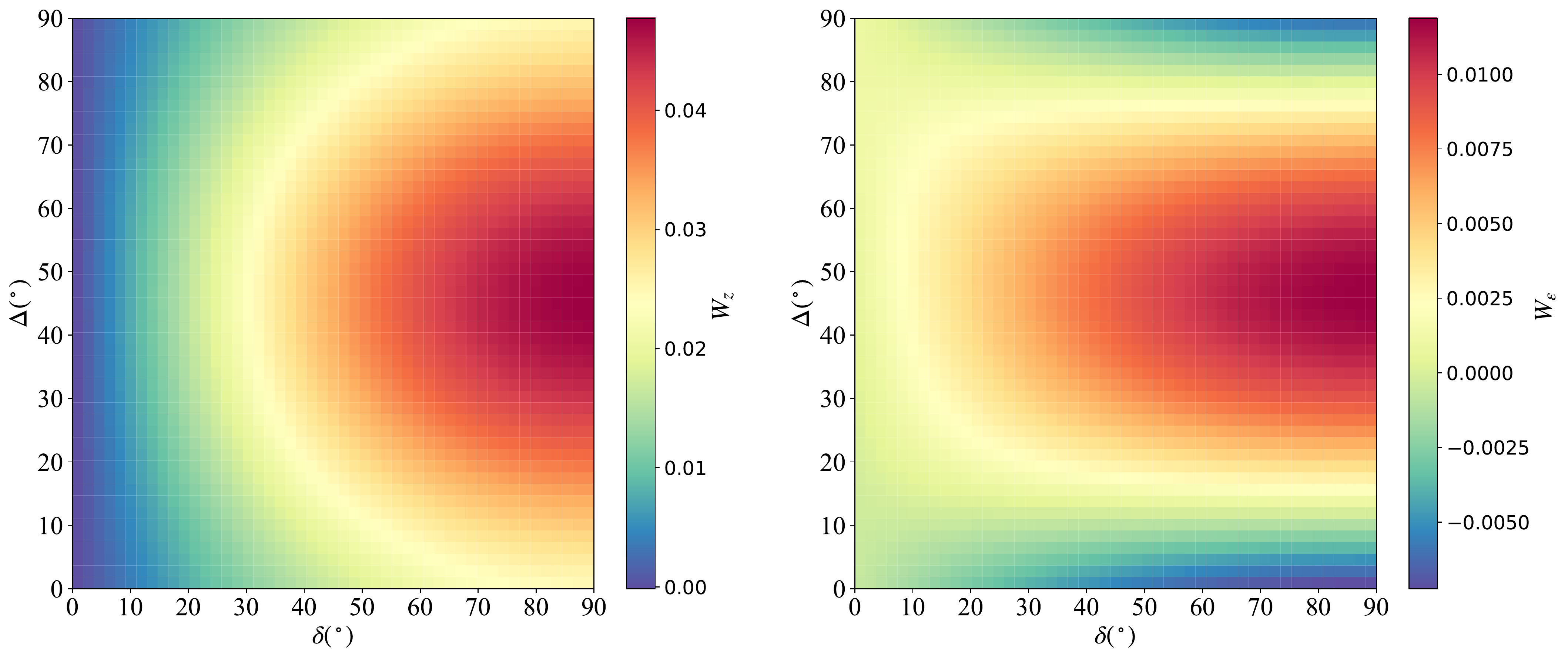}
    \caption{Map of the dimensionless parameter $W$ scaled by color from blue (low) to red (high) as a function of $\delta$ and $\Delta$ (which are related to the asymmetry of the asteroid; see Sec. \ref{sec3_1}). The spin component $W_z$ is shown in the left panel, and the obliquity component $W_\epsilon$ is shown in the right panel. Here both $\alpha$ and $\epsilon$ are set to be $\pi/4$.}
    \label{fig3}
\end{figure*}

% The first term is the torque of the tangential recoil force and the second term is that of normal recoil force. We define
% \begin{align}
% \label{eq:T_t}
%     & \vec T_{\rm t} = F_1(\lambda) \vec r_0 \times  \vec e_x \\
%     & \vec T_{\rm n} = F_3(\lambda) \vec r_0 \times  \vec e_z, 
% \end{align}
% with which we have $\vec T = \vec T_{\rm t}+\vec T_{\rm n}$.
% We will see in the Sec.~\ref{sec3} that the component $T_{\rm n}$ highly depends on the shape of the asteroid and the location of the crater while the component $T_{\rm n}$ does not. In the majority of cases, the total CYORP is dominated by $T_{\rm t}$.

In order to understand this CYORP effect on the secular spin evolution of an asteroid, we need to average it over its dynamic timescale. It is well known that the timescale of the YORP effect is much longer than the orbital period and spin period, therefore it is useful to calculate the average YORP torque over the orbital period and the spin period. In general, the spin period ($\text{some}$ hours) is much shorter than the orbital periods ($\text{some}$ years), so that the integral over the orbital motion and that over the spin motion can be treated separately. 

We consider an inertia reference frame ($XYZ$) with the origin $O$ at the asteroid center (see Fig.~\ref{fig2}). The axis $OZ$ is the spin axis of the asteroid, and the $OXY$ plane is the equatorial plane. The position vector of the crater $\vec r_0$ is 
\begin{equation}
\label{eq:r_0}
    \vec r_0 = r_0 (\sin \alpha \cos \beta, \sin \alpha \sin \beta, \cos \alpha),
\end{equation}
where $r_0$ is the distance from the crater to the mass center of the asteroid. The unit normal vector $n_0$ of the ground can be expressed as
\begin{equation}
\label{eq:n_0}
    \vec n_0 = (\sin (\alpha+\Delta) \cos (\beta+\delta), \sin (\alpha+\Delta) \sin (\beta+\delta), \cos (\alpha+\Delta))= -\vec e_z,
\end{equation}
where the independent variables $\delta$ and $\Delta$ denote deviations of the normal vector from the position vector, which is determined by the geometry of the asteroid.
The unit solar vector $\vec s$ is 
\begin{equation}
    \vec s = (\cos \eta, \cos \epsilon \sin \eta, \sin \epsilon \sin \eta),
\end{equation}
where $\eta$ is the angle of orbital motion. The vector $\vec e_x$ can be expressed as a function of $\vec s$ and $\vec e_z$ according to Equation~(\ref{eq:s}).
Therefore, the averaged CYORP torque is 
\begin{equation}
\label{eq:T_average}
    <\vec T_{\rm CYORP}> = \frac{1}{4\pi^2} \int_0^{2\pi} \int_0^{2\pi}  \vec T_{\rm CYORP} H(\vec s \cdot \vec n_0) d\beta d\eta.
\end{equation}
Here $H$ is the Heaviside step function, which is defined as
\begin{equation}
    H(x) \coloneqq 
    \left \{
    \begin{aligned}
    & 1, x>0 \\
    & 0, x\leq 0.
    \end{aligned}
    \right.
\end{equation}
The average CYORP torque turns out to be a function of $\gamma_0$, $\alpha$, $\epsilon$, $\delta$ and $\Delta$, where
\begin{description}
\item[$\gamma_0$] describes the depth-diameter ratio of the crater (see Fig.~\ref{fig1}),
\item[$\alpha$] is the latitude of the crater,
\item[$\epsilon$] is the obliquity of the asteroid,
\item[$\Delta$] and $\delta$ describes the deviation of the normal vector $\vec n_0$ from the position vector $\vec r_0$ (see Eq. \ref{eq:n_0}), which is determined by the macroscopic geometry of the asteroid. 
\end{description}
The dependence of the CYORP effect on these parameters is exposed in the next section.

\section{Results}
\label{sec3}

\begin{figure*}
    \centering
    \includegraphics[width=\textwidth]{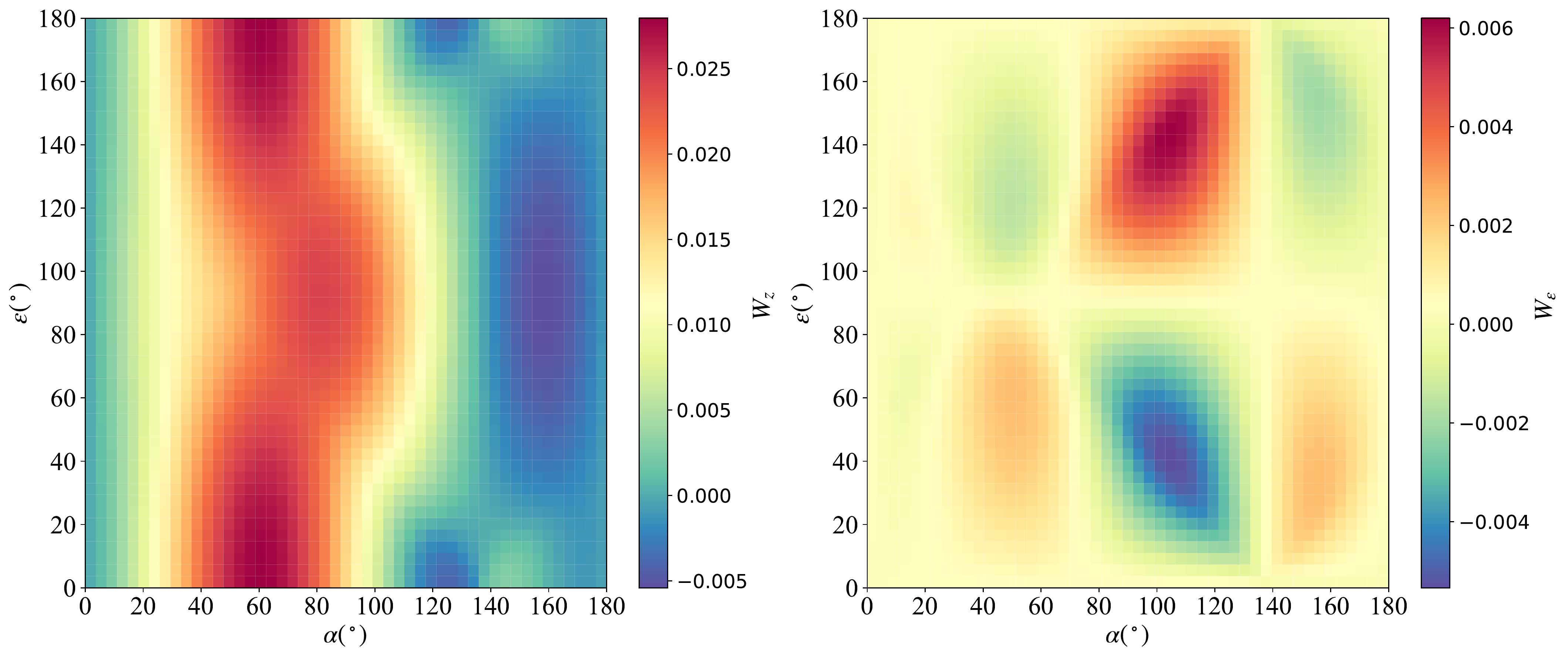}
    \caption{Map of the dimensionless parameter $W,$ which varies with the latitude of the crater $\alpha$ and the obliquity of the asteroid $\epsilon$, scaled by color from blue (low) to red (high). The spin component $W_z$  is shown in the left panel, and the obliquity component $W_\epsilon$ is shown in the right panel. Here both $\delta$ and $\Delta$ are set to be $\pi/4$.}
    \label{fig4}
\end{figure*}

In general, the averaged CYORP torque has the form
\begin{align}
\label{eq:T_CYORP_g1}
    & <\vec T_{\rm CYORP}>_z = W_z(\gamma_0, \alpha, \epsilon, \Delta, \delta) \frac{\Phi}{c}R_0^2  R_{\rm ast}, \\
\label{eq:T_CYORP_g2}
    & <\vec T_{\rm CYORP}>_\epsilon = W_\epsilon(\gamma_0, \alpha, \epsilon, \Delta, \delta) \frac{\Phi}{c}R_0^2  R_{\rm ast},
\end{align}
where the $W$ function can be obtained from Equation~(\ref{eq:T_average}). Here $<\vec T_{\rm CYORP}>_z$ and $<\vec T_{\rm CYORP}>_\epsilon$ are equivalent to the $Z$ and $Y$ components of the CYORP torque, denoting the spin and obliquity torques, respectively. In the following analysis, we present the values of $W_z$ and $W_\epsilon$ in different cases to show how the CYORP torque varies with the parameters. 

\subsection{Asteroid shape}
\label{sec3_1}
The asteroid shape affects the relation between $\vec r_0$ and $\vec n_0$. For example, a spherical asteroid has $\vec r_0/r_0 =  \vec n_0$ , while a prolate asteroid does not. The relation between $\vec r_0$ and $\vec n_0$ translates into $\Delta$ and $\delta$ (see Eqs.~(\ref{eq:r_0}) and (\ref{eq:n_0})) in our calculation. Figure \ref{fig3} shows the values of the dimensionless parameter $W$ in terms of $\delta$ and $\Delta$ , where $\alpha = \epsilon = \pi/4$. We demonstrate that for $Z$-axis symmetric asteroids, which is equivalent to $\delta = 0$, the CYORP torque disappears due to the antisymmetry of the torque function over the integral domain.
However, for nonsymmetric asteroids ($\delta \neq 0$), the CYORP torque includes both the spin and obliquity components. 

\subsubsection{$Z$-axis symmetric asteroid}
\label{sec3_1_1}
Here, we call a $Z$-axis symmetric asteroid an asteroid that has a surface of revolution around the $z$-axis (major principal axis)\footnote{Cross sections perpendicular to the $Z$-axis are circular.}. Some well-known examples are top-shape asteroids and symmetric ellipsoid asteroids with an axis ratio $1:1: c_l$ ($c_l$>0). A $Z$-axis symmetric asteroid has $\delta = 0$ everywhere on its surface, as demonstrated in Appendix \ref{appB}. Given $\delta = 0$, we substitute Equation~(\ref{eq:r_0}) and Equation~(\ref{eq:n_0}) into Equation~(\ref{eq:T_CYORP2}), leading to
\begin{equation}
\label{eq:T_CYORP3}
\begin{aligned}
     \vec T_{\rm CYORP}  & = \frac{F_1(\lambda)}{\sin \lambda}
    \left(
    \begin{aligned}
    & \sin \eta \sin \alpha \sin \beta \sin \epsilon - \sin \eta \cos \alpha \cos \epsilon \\
    & \cos \eta \cos \alpha - \sin \eta \sin \alpha \cos \beta \sin \epsilon \\
    & \sin \eta \sin \alpha \cos \beta \cos \epsilon - \cos \eta \sin \alpha \sin \beta 
    \end{aligned}
    \right)   \\
    & + \left(\frac{F_1(\lambda)}{\tan \lambda} + F_3(\lambda) + \pi R_1^2 \frac{2\Phi}{3c} \cos \lambda \right)
    \left(
    \begin{aligned}
    & \sin \beta \sin \Delta \\
    & \cos \beta \sin \Delta \\
    & 0
    \end{aligned}
    \right),
\end{aligned}
\end{equation}
% For an arbitrary $z$-axis symmetric asteroid, the position vector of a point on the surface can be expressed as
% \begin{equation}
% \label{eq:r_0_1}
%     \vec r_0 = (k \cos \zeta, k \sin \zeta, p(k)),
% \end{equation}
% where $p(k)$ is a function of $k$, depending on the specific shape of the asteroid. Then the unit normal vector is
% \begin{equation}
% \label{eq:n_0_1}
%     \vec n_0 = - \frac{ {\rm d}p /{\rm d} k}{\sqrt {1 + ({\rm d}p /{\rm d} k)^2}} (\cos \zeta, \sin \zeta, -\frac{1}{{\rm d}p /{\rm d} k} ).
% \end{equation}
% We see that both $\vec r_0$ and $\vec n_0$ share the same azimuth angle $\zeta$. Therefore, for any point on a $Z$-axis symmetric asteroid, the difference between the azimuth angles of the position vector and the normal vector is zero, which means $\delta = 0$ in Equations (\ref{eq:n_0_1}).
% Plugging Equations (\ref{eq:r_0_1}) and (\ref{eq:n_0_1}) into Equation~(\ref{eq:T_CYORP2}), we have
% \begin{equation}
% \label{eq:T_CYORP3}
%     \vec T_{\rm CYORP} = \frac{F_1(\lambda)}{\sin \lambda}
%     \left(
%     \begin{aligned}
%     & -p \sin \eta \cos \epsilon + k \sin \eta \sin \epsilon \sin \zeta, \\
%     & p \cos \eta - k \cos \zeta \sin \epsilon \sin \eta,  \\
%     & k \cos \epsilon  \cos \zeta \sin \eta -k \cos \eta \sin \zeta))   
%     \end{aligned}
%     \right)
% \end{equation}
where 
\begin{equation}
\label{eq:cos_lambda}
    \begin{aligned}
        \cos \lambda &= \vec n_0 \cdot \vec s  \\
        & = \cos \eta \sin (\alpha + \Delta) \cos \beta + \sin \eta \sin (\alpha + \Delta) \sin \beta \cos \epsilon \\
        & + \sin \eta \cos (\alpha + \Delta) \sin \epsilon.
    \end{aligned}
\end{equation}
The secular CYORP torque is calculated by averaging $\vec T_{\rm CYORP} H(\cos \lambda)$ over the spin angle $\zeta$ and orbital angle $\eta$ according to Equation~(\ref{eq:T_average}). 

\begin{figure*}
    \centering
    \includegraphics[width=\textwidth]{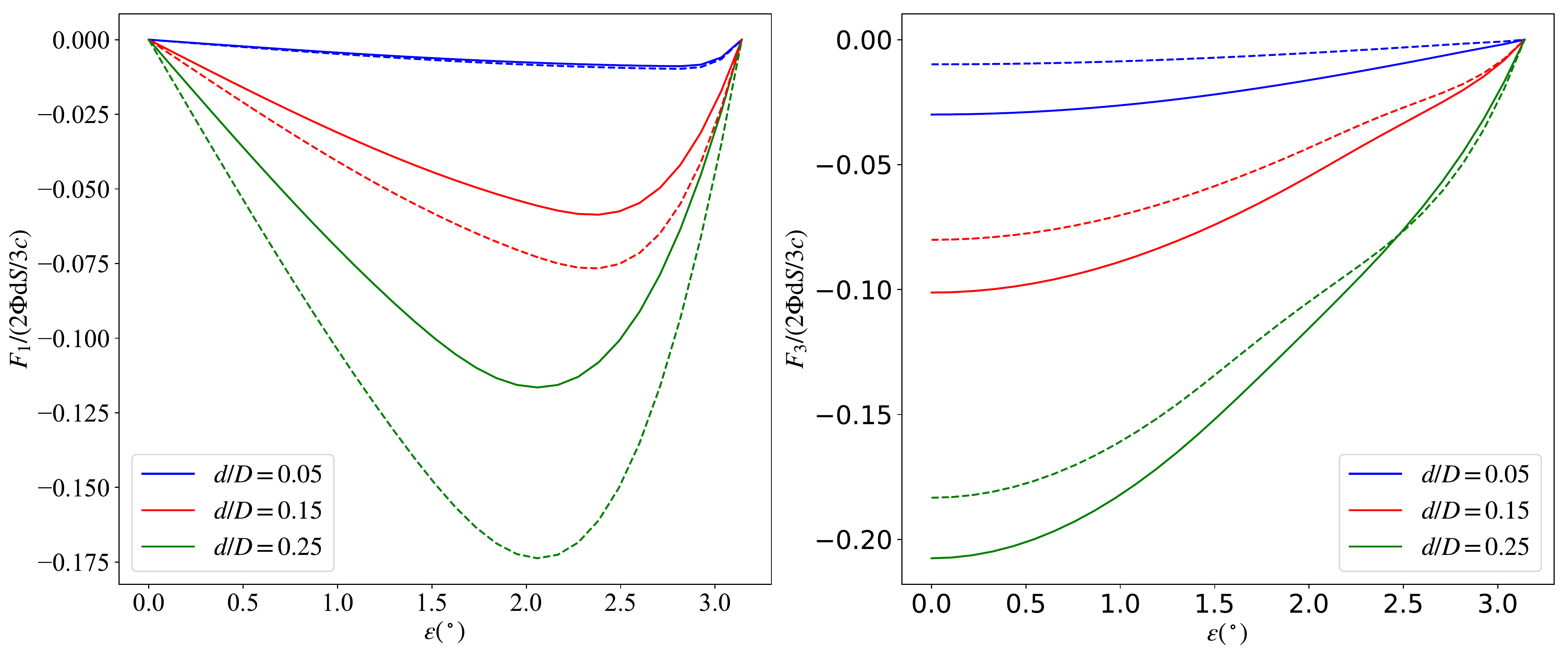}
    \caption{Recoil forces for craters with different $h/D_0$. The results considering the self-heating effect are shown by solid lines, and the results obtained without the self-heating effect are shown by dashed lines. The left panel denotes the tangential component of the recoil force, and the right panel denotes the normal component.}
    \label{fig5}
\end{figure*}

Interestingly, after averaging, the obliquity component ($Y$ component) and the spin component ($Z$ component) vanish because they are antisymmetric in the interval domain $\beta \in (0, \, 2\pi)$ and $\eta \in (0,\,2\pi)$. This becomes clear in the example of the domain $\beta \in (0, \, \pi)$ and $\eta \in (0,\,\pi)$. Because $\sin (\pi - x) = \sin x $ and $\cos(\pi - x) =- \cos x$, for any point pair $(\zeta, \eta)$, we can find that another point pair $(\pi - \zeta, \pi - \eta)$ exists for which the $Y$-axis and $Z$-axis components of $\vec T_{\rm CYORP} H(\cos \lambda)$ have the same absolute value but the opposite sign; this is shown by investigating Equation~(\ref{eq:T_CYORP3}) and Equation~(\ref{eq:cos_lambda}). The $Y$-axis and $Z$ -axis components of $\vec T_{\rm CYORP}$ change sign, but $\lambda$ does not change at all. In the domain $\zeta \in (0, \, \pi)$ and $\eta \in (0,\,\pi)$, the average function is therefore antisymmetric about $(\pi/2,\,\pi/2)$ for $Y$-axis and $Z$-axis components, which leads to the fact that the average is 0. Other antisymmetric points in the whole domain are $(\pi/2, 3\pi/2)$, $(3\pi/2, \pi/2),$ and $(3\pi/2, 3\pi/2)$ . Therefore, there is no spin and obliquity component of the CYORP torque left for $Z$-axis symmetric asteroids ($\delta = 0$). This antisymmetric property does not occur in the $X$-axis component of $<\vec T_{\rm CYORP}>$, which changes the precession angle of the asteroid. Although Nature knows no perfectly $Z$-axis symmetric asteroid, this analysis implies that the torque would be severely weakened for a nearly $Z$-axis symmetric asteroid (small $\delta$), which is also shown in Figure~\ref{fig3}. However, this antisymmetry of $T_{\rm CYORP}$ is only valid when Rubincam's approximation (zero thermal inertial) is applied and would be broken in the case of nonzero thermal inertia, for which the spin and obliquity components still exist (see Sec.~\ref{sec3_4}).

% Note that the real asteroid can hardly have $\delta$ exactly equal to zero even if it is a nearly top-shaped asteroid, so cautions must be taken when applying this rule.

% Since $\delta$ is a local quantity which is related to the vicinal geometry of the crater, we need to calculate $\delta$ in the practical case instead of simply considering whether the asteroid is a symmetric shape.
% In the next section, we will investigate in the non-perfectly-symmetric case, how $\delta$ affects the CYORP effect.

\subsubsection{Asymmetric asteroid}

A perfectly $Z$-axis symmetric asteroid does not exist in Nature, for which even the normal YORP effect vanishes \citep{Breiter2007}. We therefore investigated how the CYORP effect depends on the asymmetry of the asteroid. For an asteroid without a perfectly symmetric shape, the position vector $\vec r_0$ and the normal vector $\vec n_0$ are not always aligned in the same longitude ($\delta \neq 0$). We already know (Sec.~\ref{sec3_1_1}) that when $\delta = 0$, the CYORP torque only has the obliquity component. In this section, we investigate how $\delta$ affects the CYORP effect in the imperfectly symmetric case. We also investigate the effect of $\Delta$. For simplicity, we fixed other parameters by setting the crater shape parameter $\gamma_0 = 0.2$, the latitude $\alpha = \pi/4,$ and the obliquity $\epsilon = \pi/4$.

Figure~\ref{fig3} shows that the spin component starts from $0$ and grows with increasing $\delta$ to a magnitude comparable to the obliquity component. Thus, in the case of $\delta \neq 0$, which is more common in real craters on asteroids, the CYORP torque has a non-negligible spin component that changes the spin rate of the asteroid in the long term.

\subsection{Crater latitude $\alpha$ and asteroid obliquity $\epsilon$}
In order to determine how $W$ varies with the crater latitude and the obliquity, we need to keep other variables constant. Figure~\ref{fig4} shows the $W$ map with a crater latitude $\alpha \in (0, \pi)$ and an asteroid obliquity $\epsilon \in (0, \pi)$ when $\delta$ and $\Delta$ are set to be $\pi/4$. The latitude of large craters can cause their shape to depart from the semi-sphere model used in our study \citep{Fujiwara1993, Daly2020}. The effect of more complex geometries is left for future studies.

\subsection{Crater depth-diameter ratio}
The above results assume that the depth-diameter ratio $h/D_0$ of the crater is $\sim 0.16$, while real craters on asteroids exhibit wide ranges of this ratio. Figure~\ref{fig5} shows the recoil forces (Eq.~(\ref{eq:F})) caused by craters with different depth-diameter ratios. Moreover, for the tangential component of the recoil force, the self-sheltering effect is negligible in shallow craters (low $h/D_0$), while for the normal component, the self-heating effect cannot be ignored even in shallow craters.

\subsection{Thermal inertia}
\label{sec3_4}
The inclusion of nonzero thermal inertia increases the complexity of the problem and requires a numerical method to obtain a precise solution, which is beyond the scope of this paper. However, we can reasonably modify the total force of the crater in order to mimic the thermal lag effect due to nonzero thermal inertia. We assumed that the Sun rises from the east and sets in the west from the view of a crater on the asteroid. The west part of the crater is illuminated in the morning and the east part is illuminated in the evening. The YORP torque arises from the temperature difference between the west and east parts. However, the temperature difference in the morning should be different from that in the evening as a result of the thermal inertia. In the morning, the crater just experienced a dark night, while in the evening, the crater has been sunlit for the whole day. This means that the temperature of the crater is {\it not} symmetric in the daytime, which will induce a nonzero $y$ component of the total recoil force $\vec F$ in Equation~(\ref{eq:F}), 
\begin{equation}
    \vec F_2(\lambda) \neq 0. 
\end{equation}
Although we are currently unable to obtain the precise solution of $F_2(\lambda)$, we can at least examine whether it has an effect on the CYORP torque by simply performing the transformation $ F_1(\lambda) \rightarrow F_1(\lambda)/\sqrt 2$ and $ F_2(\lambda) \rightarrow F_1(\lambda)/\sqrt 2$. Here a hidden assumption is that $F_1(\lambda)$ and $F_2(\lambda)$ are on the same order of magnitude. This transformation does not give a direct estimate of the considered thermal inertia, and it is used here only to account for the effect of nonzero thermal inertia. In future work, we will directly estimate the consequences of given values of thermal inertia on CYORP.

\begin{figure*}
    \centering
    \includegraphics[width = \textwidth]{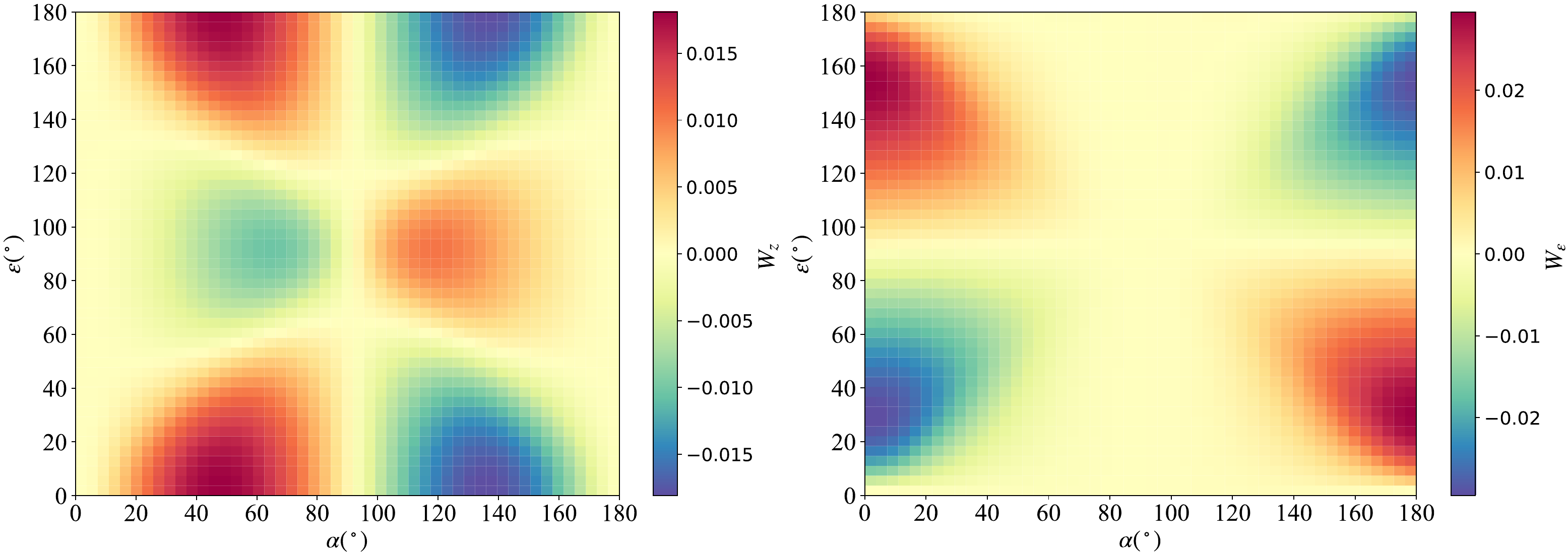}
    \caption{Color map of the dimensionless parameters $W_z$ (left panel) and $W_{\epsilon}$ (right panel) accounting for nonzero thermal inertia. The variables $\delta$ and $\Delta$ are set to zero. Here a nonzero value of $W_z$ and $W_\epsilon$ arises, while the values are zero throughout the map in the regime of zero-thermal inertia, as proven in Section~\ref{sec3_1_1}.}
    \label{fig6}
\end{figure*}

% Previous research already show that for NYORP torque, the thermal inertia does not affect the spin component, so here we can examine the spin component of the CYORP torque without introducing any inaccuracy caused by the thermal lag of $\vec T_{\rm ground}$ (the NYORP term in the CYORP, see Equation~(\ref{eq:T_CYORP1})). 
Figure~\ref{fig6} shows the values obtained for $W_z$ and $W_\epsilon$ when $\delta = \Delta = 0$. With nonzero thermal inertia, the spin and obliquity components arise for some sets of ($\alpha,\, \epsilon$), while they are always zero without thermal inertia due to $\delta = 0, $ which has already been proven in Sec.~\ref{sec3_1_1}. Therefore, we infer that the nonzero thermal inertia of the asteroid can induce a nonzero spin component of the CYORP torque for $Z$-axis symmetric asteroids, and should affect the behavior of the CYORP torque; this will be studied in future works.

\section{Discussion and implications}
\label{sec4}
\subsection{Order of magnitude}
\label{sec4_1}
In order to understand the effectiveness of the CYORP torque better, it is useful to compare it with the two torques in the current YORP model: the normal YORP (NYORP) torque, which is caused by the global asymmetry of the asteroid, and the tangential YORP (TYORP) torque, which results from the temperature difference of two sides of boulders. To a first approximation, the normal YORP torque for an asteroid can be simply expressed as  
 \begin{align}
 \label{eq:T_NYORP1}
     & <\vec T_{\rm NYORP,z}> \sim C_z \frac{\Phi}{c} R_{\rm ast}^3 \cos (2\epsilon + \frac{1}{3}), \\
\label{eq:T_NYORP2}
     & <\vec T_{\rm NYORP,\epsilon}> \sim C_{\epsilon} \frac{\Phi}{c} R_{\rm ast}^3 \sin 2\epsilon.
 \end{align}
Here $C_z$ and $C_{\epsilon}$ are dimensionless YORP coefficients of the spin component and obliquity component, respectively. \citet{Golubov2019} computed the normal YORP torques for type \uppercase\expandafter{\romannumeral1}, \uppercase\expandafter{\romannumeral2}, \uppercase\expandafter{\romannumeral3}, and \uppercase\expandafter{\romannumeral4} asteroids \footnote{Asteroids are categorized into types \uppercase\expandafter{\romannumeral1}, \uppercase\expandafter{\romannumeral2}, \uppercase\expandafter{\romannumeral3}, and \uppercase\expandafter{\romannumeral4} according to the behavior of the YORP torque curve \citep{Vokrouhlicky02}} from the sources of photometric observations, radar measurements, and in situ observations. For type \uppercase\expandafter{\romannumeral1} and \uppercase\expandafter{\romannumeral2} asteroids, the number distribution of $C_{z}$ peaks around 0.005, while for type \uppercase\expandafter{\romannumeral3} and \uppercase\expandafter{\romannumeral4} asteroids, the peak is located at $C_z < 0.001$.  Here we took $C_z = 0.005$ for the following comparison. It was shown that an approximate correlation between these two coefficients is given by $C_\epsilon/ C_z \sim 2/3$ \citep{Golubov2019, Marzari2020}. 

The tangential YORP torque for one boulder, which is dominated by the spin component, is \citep{Golubov12}
\begin{equation}
\label{eq:T_TYORPb}
    <T_{\rm TYORP,b}> = C_{\rm T,b} S  R_{\rm ast} \frac{\Phi}{c},
\end{equation}
where $S$ is the projection area of the boulder on the ground base. The parameter $C_{\rm T,b}$ measures the efficiency of the torque, depending on the thermal parameter and the shape model \citep{Golubov17}. For a spherical boulder, $C_{\rm T,b} \sim 0.002,$ while for a wall, $C_{\rm T,b} \sim 0.01$. The numerical simulation by \citet{Sevevcek2015} on a polyhedron model of the boulder found $C_{\rm T,b} \sim 0.001$. When all the boulders on asteroid Itokawa were considered, the total TYORP torque was 
\citep{Golubov2019}
\begin{equation}
     <\vec T_{\rm TYORP}> \sim C_{\rm T} \frac{\Phi}{c} R_{\rm ast}^3 \exp \left( -{(\ln \Theta - \ln \Theta_0) \over \Gamma^2} \right) (\cos^2 \epsilon + 1).
\end{equation}
For spherical boulders, $\Gamma = 1.518$ and $\ln \Theta_0 = 0.58$. The coefficient $C_{\rm T}$ depends on the roughness of the surface and on the shape of the asteroid. For asteroid (25143) Itokawa, the value of $C_{\rm T}$ was estimated as $0.0008 \pm 0.0005$ \citep{Sevevcek2015,Marzari2020}.

The general form for the CYORP torque of one crater is similar to Equation (\ref{eq:T_TYORPb}),
\begin{equation}
\label{eq:T_CYORPc}
    <T_{\rm CYORP}> = W \frac{\Phi}{c} R_0^2  R_{\rm ast}.
\end{equation}
When $\delta = \Delta = \alpha = \pi/4$, $W_z$ is about 0.04 and $W_\epsilon$ is about 0.025. Comparing Equation (\ref{eq:T_CYORPc}) to Equation (\ref{eq:T_TYORPb}), we find that the CYORP torque is one order of magnitude stronger than the TYORP torque for a crater and a boulder (spherical model) of the same size.

\begin{figure}
    \centering
    \includegraphics[width = 0.5 \textwidth]{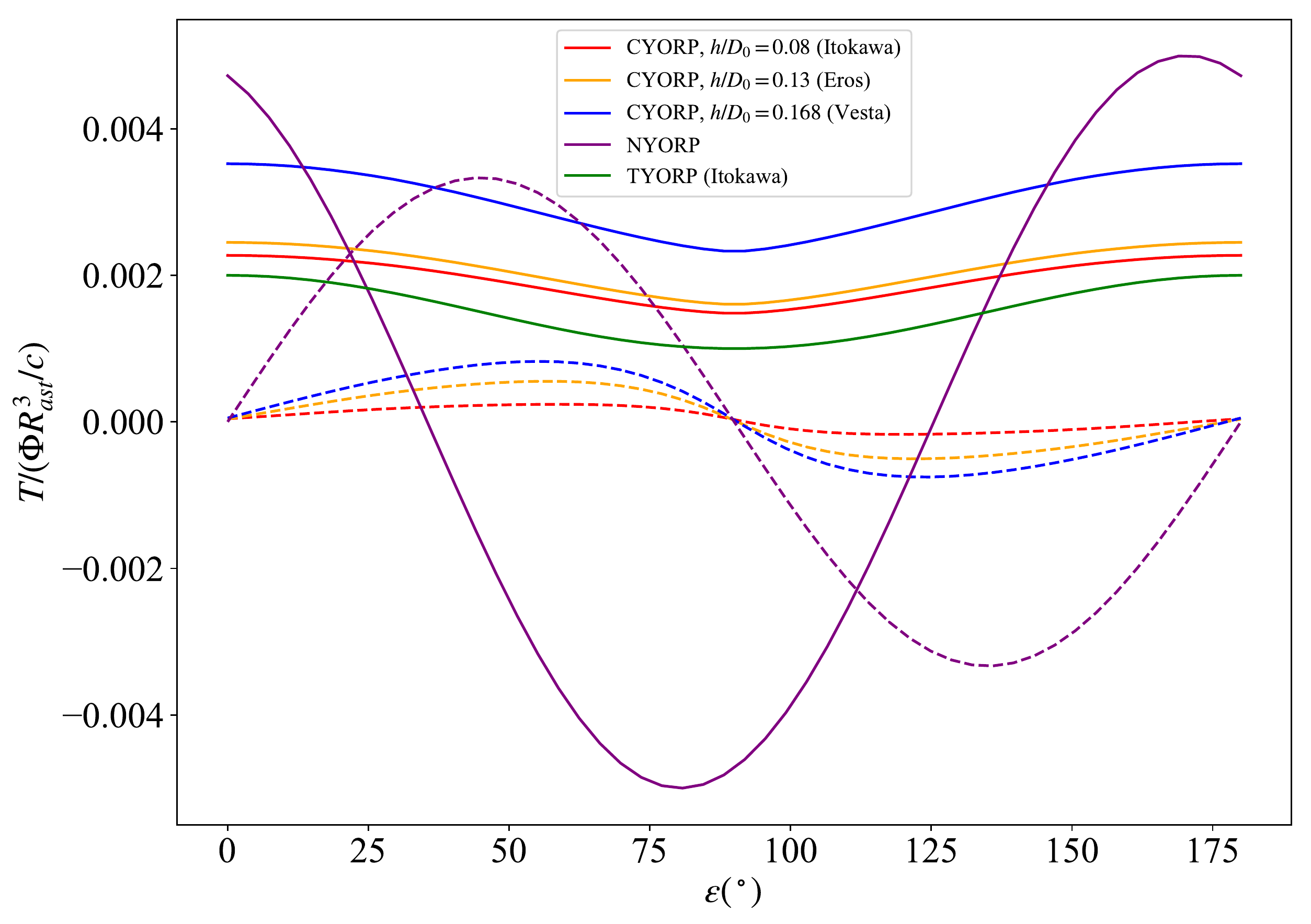}
    \caption{Comparison between the CYORP, TYORP, and NYORP torques, which are shown in different colors. The solid curve denotes the spin component, and the dashed curve denotes the obliquity component.}
    \label{fig7}
\end{figure}

In Figure~\ref{fig7} we compare the CYORP torque for a single crater to the NYORP and TYORP torques for the whole asteroid as functions of the obliquity $\epsilon$. Here parameters and the magnitude of the TYORP (e.g., the size distribution and the thermal parameter of boulders) follow the research performed on asteroid Itokawa \citep{Sevevcek2015}. The TYORP torque differs from one asteroid to the next because the morphology of asteroids differs \citep{Kanamaru2021}. For the CYORP torque, other parameters apart from the obliquity were set to be constant as $\delta = \Delta = \alpha = \pi/4$ and $R_0/ R_{\rm ast} = 1/3$. We considered three crater depth-diameter ratios, $h/D_0 $=0.08, 0.13, and 0.168, as examples, which are the mean values for asteroids Itokawa, Eros, and Vesta, respectively \citep{Hirata2009, Robinson2002, Vincent2014}. In these cases, the values of $W_z$ are 0.04, 0.028, and 0.025, respectively, while $W_\epsilon$ is much smaller. For a deep crater with $h/D_0 = 0.168$, the CYORP torque for {\it a single crater} is comparable to the NYORP torque. CYORP decreases with decreasing depth-diameter ratio, but even for a shallow crater with a depth-diameter ratio $\sim 0.08$, the CYORP torque is stronger than the total TYORP torque for the whole asteroid. 

Although we assumed a large crater with a size one-third of the size of the asteroid to calculate the CYORP torque, large craters like this exist on real asteroids \citep[e.g., asteroids Itokawa and Ryugu,][]{Hirata2009, Noguchi2021}. According to \citet{Hirata2009}, the largest three craters on Itokawa are 134 m ($h/D_0 \sim 0.11$), 128 m ($h/D_0 \sim 0.12$), and 117 m ($h/D_0 \sim 0.15$). Considering that the mean diameter of asteroid Itokawa is $\sim 313$m, the value of $R_0/ R_{\rm ast}$ set to $\text{one-third}$ is reasonable for real asteroids and appropriate for asteroid Itokawa. It might be twice the total torque or cancel the NYORP torque, depending on whether the sign of the CYORP torque is opposite to that of the TYORP torque. In Figure \ref{fig7}, the CYORP torque is positive all over the obliquity, which is not the same for all cases, however, leading to the change in sign of the total YORP torque in some obliquities when the CYORP torque is added to the NYORP torque. Therefore, we show that the CYORP effect might be the main complement to the NYORP effect in addition to the TYORP effect. Especially when the thermal inertia of the asteroid is extremely low, the TYORP effect vanishes, so that the CYORP effect might be the only complement to the NYORP effect.

% is about twice the NYORP torque, which potentially results in the change of sign of the total YORP torque at some obliquities. For shallower craters (e.g. $h/D_0 = 0.08$), the CYORP torque is a bit stronger than TYORP. TYORP vanishes when the thermal inertia is zero while CYORP does not. So when the asteroid is in a very slow rotation or has a low thermal conductivity, the CYORP torque is the major contributor to the NYORP effect.
The CYORP torque for a smaller crater decreases as the CYORP torque scales as $\sim R_0^2$, which, however, does not mean that the contribution of small craters to the total CYORP torque is negligible. On the contrary, small craters could even give rise to a more significant CYORP torque than that produced by large craters because there are many small craters. This is analyzed in more detail in the next section as this section focuses on the order of magnitude of the CYORP torque of a single crater.

% Comparing Equations (\ref{eq:T_CYORP_g1})(\ref{eq:T_CYORP_g2}) with Equations (\ref{eq:T_NYORP1})(\ref{eq:T_NYORP2}), we have
% \begin{align}
% & \frac{ <\vec T_{\rm CYORP,z}>}{ <\vec T_{\rm NYORP,z}>} \sim \frac{W_z}{C_z} \left(\frac{R_0}{R}\right)^2, \\
% & \frac{ <\vec T_{\rm CYORP,\epsilon}>}{ <\vec T_{\rm NYORP,\epsilon}>} \sim \frac{W_\epsilon}{C_\epsilon} \left(\frac{R_0}{R}\right)^2.
% \end{align}
% By setting the above two expressions equal to 1, we obtain the approximate size of the crater that would produce a torque that is comparable to the NYORP torque. a crater with the radius 1/3 of the asteroid radius could produce an {\it obliquity} torque comparable to the NYORP torque of the whole asteroid, and a crater with .
% Note this approximation is based on $W_\epsilon \sim 0.02$, while Figure~\ref{fig3}-\ref{fig5} show that the value of the $g$ function has a wide range, depending on the depth-diameter ratio, 

% \begin{equation}
%     \vec T_{\rm YORP} = \vec T_{\rm NYORP} + \vec T_{\rm TYORP},
% \end{equation}
% where $\vec T_{\rm NYORP}$ and $\vec T_{\rm TYORP}$ stand for the normal YORP effect and the tangential YORP effect, respectively. Our study shows that the CYORP might account for the total YORP torque as well, which adds a {\it CYORP} term into the above equation: 
% \begin{equation}
%     \vec T_{\rm YORP} = \vec T_{\rm NYORP} + \vec T_{\rm TYORP} + \vec T_{\rm CYORP},
% \end{equation}
% with 
% \begin{equation}
%     \vec T_{\rm CYORP} = \Sigma_i T_{\rm CYORP,i}
% \end{equation}
% as a summation for a whole set of craters or concave structures on the asteroid.

\subsection{Applicability}
\label{sec4_2}
The CYORP effect, which is induced by concave structures on an asteroid surface, is expected to have widespread applications in the rotational dynamics of asteroids. It contributes to the accurate calculation of the complete YORP torque by providing a systematical assessment of the YORP torques from large-scale (craters) and small-scale (roughness) concave structures over a huge parameter space (Sec. \ref{sec4_2_1}). In addition, the CYORP effect is linked to the collision history of asteroids, whose surfaces are modified by impact craters. The spin rate and obliquity are expected to go through a random walk under the CYORP effect. This has strong implications on the rotational and orbital evolutions of asteroids (Sec. \ref{sec4_2_2}).
% Although we are at a early stage of understanding the CYORP effect, we still see

% This provides important implications in the measurement of the complete YORP torque and also brings new insights to our understanding of the spin evolution of asteroids  
                    
\subsubsection{Calculation of the total YORP torque}
\label{sec4_2_1}

Recently, in situ observations by spacecraft provided high-resolution images of asteroids and measurements of their physical properties \citep[e.g.,][]{Hirata2009, Daly2020b}, which enabled investigating the YORP effect with a high-resolution shape model of the considered asteroid \citep{Kanamaru2021, Roberts2021}. With high-resolution images, the thermally induced torque of craters whose sizes are above the image resolutions can be well represented by the NYORP torque, but we still miss the consideration of small craters or concave structures that are below the image resolution.

As we highlight in Section \ref{sec4_1}, the CYORP torque produced by small craters and not by large craters might be the main contributor to the total CYORP effect. The cumulative size distribution of craters is typically represented by a power law of the form
\begin{equation}
    N(R \ge R_0) \propto R_0^{-p}
\end{equation}
\citep[e.g., $p \sim 3$ for asteroid Itokawa;][]{Hirata2009}. The total CYORP torque of craters is simply the sum of torques due to all the craters (see Eq. (\ref{eq:T_CYORP_total})), which reads
\begin{equation}
\label{eq:T_CYORP_total2}
    \vec T_{\rm CYORP,total} \geq  N(R\geq R_0) \cdot W\frac{\Phi}{c}R_0^2  R_{\rm ast} \propto W\frac{\Phi}{c}R_0^{2-p}  R_{\rm ast}.
\end{equation}
In the case of $p>2$, the total CYORP torque is dominated by small craters because it scales as $R_0^{2-p}$. Some of the CYORP torques may cancel out due to the opposite signs of the torques over different latitudes. However, because the torque curve is not antisymmetric over the latitude (see Fig.~\ref{fig4}), it is still possible that there are enough torques of one sign to keep the net value of the torque from all craters on the same order of magnitude as the equivalent torque from a single large crater. Essentially, the CYORP torque depends on the total area of the craters on the surface. For example, 100 craters, each covering an area of $1~\rm m^2$, are equivalent for the CYORP torque to one single crater covering an area of $100~\rm m^2$. 

The concept of a "crater" in this paper can be extended to any concave structure as we do not use other properties of craters than the shape. It is hard to confine a characteristic size range of the CYORP effect because it works for all sizes in principle. Therefore, even though the resolution of the shape model might seem high from in situ observations \citep[e.g., $80~\rm cm$ in the shape model of Bennu;][]{Barnouin2019}, it may still not be high enough to resolve the small surface structures that could nevertheless induce a considerably strong CYORP torque. This poses challenges to the precise measurement of the total YORP torque. The definition of a very small crater is vague, and to compute YORP torques, the term "roughness" may be more appropriate. In this sense, the CYORP effect resulting from small-scale concave structures serves the same purpose as the YORP effect from surface roughness \citep{Rozitis2012}. \citet{Rozitis2012} showed with numerical simulations that the surface roughness mainly dampens the total YORP effect, while our result based on a semi-analytical method shows that the CYORP effect may either enhance or weaken the YORP torque, depending on many factors (see Fig \ref{fig3}, \ref{fig4}, and \ref{fig6}). Furthermore, by using a semi-analytical method, which is much faster to run over the whole parameter space, we performed a systematic investigation of how the CYORP effect depends on the properties of the craters and the asteroid. The CYORP effect resulting from the roughness on a surface provides a potential explanation for the inconsistency of the YORP model that have been encountered so far with the measurement in the case of asteroid Itokawa, even though high-resolution shape models were used  \citep{Vokrouhlicky2004, Scheeres2007, Lowry2014, Breiter2009, Sevevcek2015}. Therefore, applying the CYORP effect to the measurement of the YORP torque caused by the surface roughness, together with the TYORP effect \citep{Golubov2022}, would be an effective way to improve the estimated accuracy of the YORP effect.
In addition, the CYORP effect can also be applied to estimating the YORP torque when it is too expensive to compute the total YORP torque on a precise shape model or when detailed information of an asteroid is unavailable.

The resolution of the OSIRIS-REx mission is $\sim 80~\rm cm$ \citep[see][]{Daly2020b}. This leads to a shape model with more than three million facets, which makes it computationally demanding and time-consuming to calculate the complete YORP torque, however. It is impractical to apply such a complicated shape model to an analysis of the YORP effect under different rotational and orbital conditions (e.g., for building a statistical database of the YORP torques). The computational expense increases the difficulty of fully investigating the rotational evolution of a particular asteroid or of an asteroid family. Moreover, precise shape models are not available for most asteroids, of which high-resolution images are lacking. Therefore, a simplified but still accurate YORP model is needed, to which the CYORP effect might contribute. 
The CYORP torque, together with the TYORP torque, might be interpreted as estimation errors to the NYORP torque through a lack of necessary information on the asteroid,
\begin{equation}
    \vec T_{\rm YORP} = T_{\rm NYORP} + \sigma_{\rm TYORP} + \sigma_{\rm CYORP}.
\end{equation}
Here $\sigma_{\rm TYORP}$ and $\sigma_{\rm CYORP}$ are the uncertainties caused by all boulders and craters, respectively. Although we do not fully understand the precise magnitude of $\sigma_{\rm TYORP}$ and $\sigma_{\rm CYORP}$ at the current stage, with more information on asteroid surfaces (e.g., the size distribution of boulders and craters) and further explorations of TYORP and CYORP effects, we would be able to estimate the YORP torque from limited information (e.g., a low-resolution shape model derived from photometric observations). 

The existence of $\sigma_{\rm CYORP}$ might explain the different distributions of the YORP torques from photometric shape models and from radar shape models that were found by previous simulations \citep{Marzari2020}. Because shape models derived from the photometry are usually convex, which means that the information on craters is lost, while those from radar data could be concave, the different distribution of the YORP torques in these two groups might be caused by the CYORP effect in the second group. In addition, to simulate the YORP effect on synthetic pseudo-asteroid shape models, shape models including concave structures \citep[e.g.,][]{Devogele2015} would be more appropriate to account for the CYORP effect.

% Therefore we suggest that future research on YORP model includes CYORP as a contributor to the total YORP effect. In the future we will apply the CYORP model to compute its effect for real asteroids, accounting for the actual size and depth distributions of craters on their surfaces. The CYORP torque can be computed for craters but also for all concave structures on the surface of an asteroid, although a modification is needed to account for the corresponding geometry of the structure.

\subsubsection{Influence on spin evolution}
\label{sec4_2_2}

The CYORP torque measures the torque difference of a crater and the ground before the occurrence of the crater (see Eq. (\ref{eq:T_CYORP1})). As the crater is produced by a collision event, the CYORP torque naturally computes the change in YORP torque before and after a collision by its definition. Each collision event produces a CYORP torque, resulting in a random walk of the YORP torque over the collisional history, which affects the spin evolution of the asteroid. Therefore, by applying the CYORP effect to an asteroid, we might be able to trace back its spin evolution assuming the crater age is known, although other factors (e.g., the boulder distribution) should be considered together. In this way, the CYORP effect builds a bridge between spin evolution and collisional history. 

We consider a crater with a radius $R_0$ on an asteroid with a radius $R_{\rm ast}$. The size of the impactor has a relation to the crater size
\begin{equation}
    R_{\rm imp} =  R_0/f_C,
\end{equation}
where the factor $f_C$ is determined by a crater scaling law \citep{Holsapple1993, Bottke2020}. The timescale of such an impact is 
\begin{equation}
    \tau_{\rm imp} = \frac{1}{P_{\rm i} \pi R_{\rm ast}^2 N(R>R_{\rm imp})}.
\end{equation}
Here $P_{\rm i} = 2.85 \times 10^{-18} \rm km^{-2} yr^{-1}$ is the intrinsic collision probability. The number of impactors that are larger than $R_{\rm imp}$ can be calculated by a simple power law \citep{Holsapple2022},
\begin{equation}
    N(R > R_{\rm imp}) = C_R \left( \frac{R_{\rm imp}}{1~{\rm km}}  \right)^{-b_R}
\end{equation}
with $C_R = 6 \times 10^5$ and $b_R = 2.2$. When the total area of craters reaches one-tenth of the asteroid surface area (equivalent to a crater with the size one-third of the asteroid radius), the YORP torque is reset by the CYORP torque, as shown in Section \ref{sec4_1}. Therefore, the critical number of impacts that can reset the YORP torque is 
\begin{equation}
    N_0 \simeq R_{\rm ast}^2/10R_0^2.
\end{equation}
Now we are able to derive the timescale for a reset of the YORP torque,
\begin{equation}
\label{eq:tau_YORP_reset1}
\begin{aligned}
    \tau_{\rm YORP,reset} &= N_{\rm 0} \tau_{\rm imp} \\
    &= \frac{R_{\rm ast}^2}{10R_{0}^2} \frac{( R_0/f_C 1~{\rm km})^{b_R}}{C_R P_i \pi R_{\rm ast}^2} \\
    &= \frac{1}{10\pi C_R P_i f_C^{b_R}} \frac{R_0^{b_R-2}}{(1 \rm km)^{b_R}}.
\end{aligned}
\end{equation}
The dependence of the $\tau_{\rm YORP,reset}$ on the crater size $R_0$ is weak, and the YORP reset effect does not depend on the asteroid size $R_{\rm ast}$. When we substitute $f_C \sim 100$ as was found from the small carry-on impactor (SCI) experiment of the mission Hayabusa2 on asteroid Ryugu \citep{Arakawa2020}, we obtain
\begin{equation}
\label{eq:tau_YORP_reset2}
    \tau_{\rm YORP,reset} \sim 0.4 \left( \frac{f_C}{100}\right)^{-b_R} \left(\frac{R_0}{100~{\rm m}}\right)^{b_R-2} \rm Myr.
\end{equation}
 Equation (\ref{eq:tau_YORP_reset2}) is a rough estimate because the factor $f_C$ and the power index $b_R$ should be functions of the asteroid size. 
The timescale ($\sim 0.4~\rm Myr$) for reorientation caused by the CYORP effect is much shorter than the typical timescale of spin axis reorientation by collisions \citep[$\sim 1~$Gyr for a 1~km radius object; see][]{Athanasopoulos2022}. This indicates that the CYORP effect may play an important role in the spin evolution of a single asteroid or asteroid families, while in  current models, the change of the YORP torque caused by a collision event, as well as other resurfacing activities \citep[e.g., regolith movement][]{Miyamoto2007, Cheng2021}, is lacking \citep{Marzari2011, Holsapple2020, Holsapple2022}. 

In addition, the CYORP effect could be applied to rotational disruption models \citep{Jacobson2014} to determine the lifetime of an asteroid; here a more comprehensive knowledge of the outcomes of rotational failure is also needed \citep{Zhang18}. Furthermore, the Yarkovsky effect, which is a radiative force that slowly changes the orbits of asteroids  \citep{Vokrouhlicky1998, Vokrouhlicky2000, Bottke2001}, highly depends on the spin obliquity, which can be altered by the CYORP torque. Thus, the CYORP effect could play an important role in the long-term orbital dynamics of asteroid families, and it might, for example, modify the V-shape evolution of asteroid families \citep{Vokrouhlicky2006, Nesvorny2015, Delbo2017, Bolin2018}.

Therefore, the CYORP effect is a mechanism that is crucial for understanding the spin evolution and even the orbital evolution of asteroids. To include the CYORP effect in a Monte Carlo simulation of the spin evolution and the orbital evolution of asteroids, we need a complete sample of all possible outcomes of the CYORP torque, which depends on the properties of craters and asteroids. At the current stage, this would not be possible because we ignored the effects of thermal inertia and of secondary illumination of the crater here, which may also be important and will be investigated in the next work.

\section{Conclusions}
\label{sec5}
We first proposed and examined the significance of the crater-induced YORP (CYORP) torque by developing a semi-analytical method. This method speeds up the computation and allows us to study the functional dependence of the CYORP on the properties of the crater and the asteroid.

CYORP arises from the torque difference produced by a crater and the ground without the crater.
The assumption of zero thermal conductivity (Rubincam's approximation) and a simple semi-sphere model of craters were implemented. {We find that the CYORP torque includes the spin and obliquity components, the values of which depend on the diameter-depth ratio, latitude and normal vector of the crater, and the obliquity and thermal inertia of the asteroid.}

We gave a general form of the CYORP torque as $\vec T_{\rm CYORP} \sim W \Phi R_0^2  R_{\rm ast}/c$ (see Eqs.~(\ref{eq:T_CYORP_g1}) and (\ref{eq:T_CYORP_g2})) and estimated the typical value of the dimensionless CYORP coefficient $W_z \sim 0.04$, $W_\epsilon \sim 0.01$ for a deep crater and $W_z \sim 0.025$, $W_\epsilon \sim 0.005$ for a shallow crater. We showed that the CYORP torque is one order of magnitude stronger than the TYORP torque for a crater and a boulder of the same size. A crater with a radius of one-third of the asteroid radius (as found on asteroid Itokawa) will produce a CYORP torque that is comparable to the NYORP torque and stronger than the TYORP torque for the whole asteroid. Craters or roughness that cover one-thenth of the asteroid surface have the same effect.
Unlike the necessary presence of thermal inertia for a nonzero value of the TYORP torque, CYORP exists without thermal inertia, which implies that for fast-spinning asteroids or asteroids with low thermal conductivity, the YORP effect will be dominated by NYORP and CYORP effects. 

Although CYORP decreases with the size of the crater as $R_0^2$, the large number of small craters may mean that the CYORP torques that are due to all small craters are non-negligible. It is the total area covered by concave structures that matters, which implies that the CYORP effect caused by the surface roughness would be crucial for the complete YORP torque (see Sec. \ref{sec4_2_1}). Previous research demonstrated the YORP sensitivity to surface roughness by a numerical method \citep{Rozitis2012}. Our work performed a systematic investigation of the YORP torque of the concave structure, which could be applied in surface roughness, over a much larger parameter space by a semi-analytical method. This lies the foundation for the accurate prediction of the YORP torque on a real asteroid. The CYORP effect provides a potential reason why the modeled YORP torque  so far was unable to match the measured value in the case of asteroid Itokawa \citep{Breiter2009} even though high-resolution shape models were applied. The CYORP effect might also explain the difference of the YORP torques between photometric-shape models, which is convex, and radar-shape models, which contain concave structure \citep{Marzari2020}. However, at the current stage, it is unclear whether the CYORP torque is dominated by large concave structures (e.g., craters) or small ones (e.g., roughness).

% One special case that we discussed is the $Z$-axis symmetric asteroid, which corresponds to, e.g., top-shape, oblate or prolate asteroids with axis ratio of $1:1:c_l$). Without thermal inertia, the CYORP torque vanishes due to the symmetry. However, the existence of thermal inertia would break up the symmetry and result in a non-zero CYORP torque. 
Moreover, because an asteroid experiences numerous impacts that lead to the production of craters during its evolution \citep{Bottke2020}, the resulting CYORP torques may cause a random walk of the spin rate and obliquity of the asteroid, which may either slow down or even prevent the YORP spin-up from occuring, deferring the formation of top shapes and binary systems based on this process \citep{WalshJacobson2015}. Our estimation showed that the timescale for reorientation of an asteroid caused by the CYORP effect is $\sim$0.4~Myr with a weak dependence on the asteroid size (see Sect. \ref{sec4_2_2}), which is much shorter than the timescale caused by collisions. This is a rough estimate, and a more complete CYORP model with nonzero thermal inertia and the secondary illumination effect is needed to build a statistic sample pool covering all possible outcomes of the CYORP torque under different conditions. The CYORP effect can have strong implications on the spin evolution and also on the orbital evolution (through the Yarkovsky effect), which will be assessed in a future work. Overall, we suggest that the CYORP effect should be incorporated into future research of the YORP effect.

\begin{acknowledgements}
      We acknowledge support from the Universit\'e C\^ote d'Azur. Wen-Han Zhou would like to acknowledge the funding support from the Origin Space Company and the Chinese Scholarship Council (No.\ 202110320014). Xiaoran Yan acknowledges the funding support from the Chinese Scholarship Council (No.\ 202006210258). Patrick Michel acknowledges funding support from the French space agency CNES and from the European Union's Horizon 2020 research and innovation program under grant agreement No.\ 870377 (project NEO-MAPP). We are thankful to Bin Ren, Liangliang Yu, Masanori Kanamaru and Yue Wang for useful discussions. 
\end{acknowledgements}

\bibliographystyle{aa} % style aa.bst
\bibliography{references}

\begin{thebibliography}{59}
\expandafter\ifx\csname natexlab\endcsname\relax\def\natexlab#1{#1}\fi

\bibitem[{Arakawa {et~al.}(2020)Arakawa, Saiki, Wada, Ogawa, Kadono, Shirai,
  Sawada, Ishibashi, Honda, Sakatani, {et~al.}}]{Arakawa2020}
Arakawa, M., Saiki, T., Wada, K., {et~al.} 2020, Science, 368, 67

\bibitem[{Athanasopoulos {et~al.}(2022)Athanasopoulos, Hanu{\v{s}}, Avdellidou,
  Bonamico, Delbo, Conjat, Ferrero, Gazeas, Rivet, Sioulas,
  {et~al.}}]{Athanasopoulos2022}
Athanasopoulos, D., Hanu{\v{s}}, J., Avdellidou, C., {et~al.} 2022

\bibitem[{Barnouin {et~al.}(2019)Barnouin, Daly, Palmer, Gaskell, Weirich,
  Johnson, Al~Asad, Roberts, Perry, Susorney, {et~al.}}]{Barnouin2019}
Barnouin, O., Daly, M., Palmer, E., {et~al.} 2019, Nature geoscience, 12, 247

\bibitem[{{Bierhaus} {et~al.}(2022){Bierhaus}, {Trang}, {Daly}, {Bennett},
  {Barnouin}, {Walsh}, {Ballouz}, {Bottke}, {Burke}, {Perry}, {Jawin}, \&
  {Lauretta}}]{Bierhaus2022}
{Bierhaus}, E.~B., {Trang}, D., {Daly}, R.~T., {et~al.} 2022, in LPI
  Contributions, Vol. 2678, LPI Contributions, 1317

\bibitem[{Bolin {et~al.}(2018)Bolin, Morbidelli, \& Walsh}]{Bolin2018}
Bolin, B.~T., Morbidelli, A., \& Walsh, K.~J. 2018, Astronomy \& Astrophysics,
  611, A82

\bibitem[{{Bottke} {et~al.}(2020){Bottke}, {Vokrouhlick{\'y}}, {Ballouz},
  {Barnouin}, {Connolly}, {Elder}, {Marchi}, {McCoy}, {Michel}, {Nolan},
  {Rizk}, {Scheeres}, {Schwartz}, {Walsh}, \& {Lauretta}}]{Bottke2020}
{Bottke}, W.~F., {Vokrouhlick{\'y}}, D., {Ballouz}, R.~L., {et~al.} 2020, \aj,
  160, 14

\bibitem[{Bottke~Jr {et~al.}(2001)Bottke~Jr, Vokrouhlicky, Broz, Nesvorny, \&
  Morbidelli}]{Bottke2001}
Bottke~Jr, W.~F., Vokrouhlicky, D., Broz, M., Nesvorny, D., \& Morbidelli, A.
  2001, Science, 294, 1693

\bibitem[{Bottke~Jr {et~al.}(2006)Bottke~Jr, Vokrouhlick{\`y}, Rubincam, \&
  Nesvorn{\`y}}]{Bottke06}
Bottke~Jr, W.~F., Vokrouhlick{\`y}, D., Rubincam, D.~P., \& Nesvorn{\`y}, D.
  2006, Annu. Rev. Earth Planet. Sci., 34, 157

\bibitem[{Breiter {et~al.}(2009)Breiter, Bartczak, Czekaj, Oczujda, \&
  Vokrouhlick{\`y}}]{Breiter2009}
Breiter, S., Bartczak, P., Czekaj, M., Oczujda, B., \& Vokrouhlick{\`y}, D.
  2009, Astronomy \& Astrophysics, 507, 1073

\bibitem[{Breiter {et~al.}(2007)Breiter, Michalska, Vokrouhlick{\`y}, \&
  Borczyk}]{Breiter2007}
Breiter, S., Michalska, H., Vokrouhlick{\`y}, D., \& Borczyk, W. 2007,
  Astronomy \& Astrophysics, 471, 345

\bibitem[{{\v{C}}apek \& Vokrouhlick{\`y}(2004)}]{Capek2004}
{\v{C}}apek, D. \& Vokrouhlick{\`y}, D. 2004, Icarus, 172, 526

\bibitem[{Cheng {et~al.}(2021)Cheng, Yu, Asphaug, Michel, Richardson,
  Hirabayashi, Yoshikawa, \& Baoyin}]{Cheng2021}
Cheng, B., Yu, Y., Asphaug, E., {et~al.} 2021, Nature Astronomy, 5, 134

\bibitem[{Daly {et~al.}(2020)Daly, Barnouin, Seabrook, Roberts, Dickinson,
  Walsh, Jawin, Palmer, Gaskell, Weirich, {et~al.}}]{Daly2020b}
Daly, M., Barnouin, O., Seabrook, J., {et~al.} 2020, Science Advances, 6,
  eabd3649

\bibitem[{{Daly} {et~al.}(2022){Daly}, {Barnouin}, {Bierhaus}, {Daly},
  {Seabrook}, {Ballouz}, {Nair}, {Espiritu}, {Jawin}, {Trang}, {DellaGuistina},
  {Burke}, {Brodbeck}, \& {Walsh}}]{Daly2022}
{Daly}, R.~T., {Barnouin}, O.~S., {Bierhaus}, E.~B., {et~al.} 2022, \icarus,
  384, 115058

\bibitem[{{Daly} {et~al.}(2020){Daly}, {Bierhaus}, {Barnouin}, {Daly},
  {Seabrook}, {Roberts}, {Ernst}, {Perry}, {Nair}, {Espiritu}, {Palmer},
  {Gaskell}, {Weirich}, {Susorney}, {Johnson}, {Walsh}, {Nolan}, {Jawin},
  {Michel}, {Trang}, \& {Lauretta}}]{Daly2020}
{Daly}, R.~T., {Bierhaus}, E.~B., {Barnouin}, O.~S., {et~al.} 2020, \grl, 47,
  e89672

\bibitem[{Delbo’ {et~al.}(2017)Delbo’, Walsh, Bolin, Avdellidou, \&
  Morbidelli}]{Delbo2017}
Delbo’, M., Walsh, K., Bolin, B., Avdellidou, C., \& Morbidelli, A. 2017,
  Science, 357, 1026

\bibitem[{Devogele {et~al.}(2015)Devogele, Rivet, Tanga, Bendjoya, Surdej,
  Bartczak, \& Hanus}]{Devogele2015}
Devogele, M., Rivet, J.-P., Tanga, P., {et~al.} 2015, Monthly Notices of the
  Royal Astronomical Society, 453, 2232

\bibitem[{{\v{D}}urech {et~al.}(2018){\v{D}}urech, Vokrouhlick{\`y}, Pravec,
  Hanu{\v{s}}, Farnocchia, Krugly, Inasaridze, Ayvazian, Fatka, Chiorny,
  {et~al.}}]{Vdurech18}
{\v{D}}urech, J., Vokrouhlick{\`y}, D., Pravec, P., {et~al.} 2018, Astronomy \&
  Astrophysics, 609, A86

\bibitem[{{Fujiwara} {et~al.}(1993){Fujiwara}, {Kadono}, \&
  {Nakamura}}]{Fujiwara1993}
{Fujiwara}, A., {Kadono}, T., \& {Nakamura}, A. 1993, \icarus, 105, 345

\bibitem[{Golubov(2017)}]{Golubov17}
Golubov, O. 2017, The Astronomical Journal, 154, 238

\bibitem[{Golubov \& Krugly(2012)}]{Golubov12}
Golubov, O. \& Krugly, Y.~N. 2012, The Astrophysical Journal Letters, 752, L11

\bibitem[{Golubov \& Lipatova(2022)}]{Golubov2022}
Golubov, O. \& Lipatova, V. 2022, arXiv preprint arXiv:2203.15567

\bibitem[{Golubov {et~al.}(2014)Golubov, Scheeres, \& Krugly}]{Golubov14}
Golubov, O., Scheeres, D., \& Krugly, Y.~N. 2014, The Astrophysical Journal,
  794, 22

\bibitem[{Golubov \& Scheeres(2019)}]{Golubov2019}
Golubov, O. \& Scheeres, D.~J. 2019, The Astronomical Journal, 157, 105

\bibitem[{Hirata {et~al.}(2009)Hirata, Barnouin-Jha, Honda, Nakamura, Miyamoto,
  Sasaki, Demura, Nakamura, Michikami, Gaskell, {et~al.}}]{Hirata2009}
Hirata, N., Barnouin-Jha, O.~S., Honda, C., {et~al.} 2009, Icarus, 200, 486

\bibitem[{Holsapple {et~al.}(2020)}]{Holsapple2020}
Holsapple, K. {et~al.} 2020, arXiv preprint arXiv:2012.15300

\bibitem[{Holsapple(1993)}]{Holsapple1993}
Holsapple, K.~A. 1993, Annual review of earth and planetary sciences, 21, 333

\bibitem[{Holsapple(2022)}]{Holsapple2022}
Holsapple, K.~A. 2022, Planetary and Space Science, 219, 105529

\bibitem[{Jacobson {et~al.}(2014)Jacobson, Marzari, Rossi, Scheeres, \&
  Davis}]{Jacobson2014}
Jacobson, S.~A., Marzari, F., Rossi, A., Scheeres, D.~J., \& Davis, D.~R. 2014,
  Monthly Notices of the Royal Astronomical Society: Letters, 439, L95

\bibitem[{Kanamaru {et~al.}(2021)Kanamaru, Sasaki, Morota, Cho, Tatsumi,
  Hirabayashi, Hirata, Senshu, Shimaki, Sakatani, {et~al.}}]{Kanamaru2021}
Kanamaru, M., Sasaki, S., Morota, T., {et~al.} 2021, Journal of Geophysical
  Research: Planets, 126, e2021JE006863

\bibitem[{Lowry {et~al.}(2014)Lowry, Weissman, Duddy, Rozitis, Fitzsimmons,
  Green, Hicks, Snodgrass, Wolters, Chesley, {et~al.}}]{Lowry2014}
Lowry, S., Weissman, P., Duddy, S., {et~al.} 2014, Astronomy \& Astrophysics,
  562, A48

\bibitem[{{Marchi} {et~al.}(2015){Marchi}, {Chapman}, {Barnouin}, {Richardson},
  \& {Vincent}}]{Marchi2015}
{Marchi}, S., {Chapman}, C.~R., {Barnouin}, O.~S., {Richardson}, J.~E., \&
  {Vincent}, J.~B. 2015, in Asteroids IV (University of Arizona Press),
  725--744

\bibitem[{Marzari {et~al.}(2020)Marzari, Rossi, Golubov, \&
  Scheeres}]{Marzari2020}
Marzari, F., Rossi, A., Golubov, O., \& Scheeres, D.~J. 2020, The Astronomical
  Journal, 160, 128

\bibitem[{Marzari {et~al.}(2011)Marzari, Rossi, \& Scheeres}]{Marzari2011}
Marzari, F., Rossi, A., \& Scheeres, D.~J. 2011, Icarus, 214, 622

\bibitem[{{Michel} {et~al.}(2020){Michel}, {Ballouz}, {Barnouin}, {Jutzi},
  {Walsh}, {May}, {Manzoni}, {Richardson}, {Schwartz}, {Sugita}, {Watanabe},
  {Miyamoto}, {Hirabayashi}, {Bottke}, {Connolly}, {Yoshikawa}, \&
  {Lauretta}}]{Michel2020}
{Michel}, P., {Ballouz}, R.~L., {Barnouin}, O.~S., {et~al.} 2020, Nature
  Communications, 11, 2655

\bibitem[{Miyamoto {et~al.}(2007)Miyamoto, Yano, Scheeres, Abe, Barnouin-Jha,
  Cheng, Demura, Gaskell, Hirata, Ishiguro, {et~al.}}]{Miyamoto2007}
Miyamoto, H., Yano, H., Scheeres, D.~J., {et~al.} 2007, Science, 316, 1011

\bibitem[{Nesvorn{\`y} {et~al.}(2015)Nesvorn{\`y}, Bro{\v{z}}, Carruba,
  {et~al.}}]{Nesvorny2015}
Nesvorn{\`y}, D., Bro{\v{z}}, M., Carruba, V., {et~al.} 2015, Asteroids IV, 29,
  7

\bibitem[{Nesvorn{\`y} \& Vokrouhlick{\`y}(2008)}]{Nesvorny2008}
Nesvorn{\`y}, D. \& Vokrouhlick{\`y}, D. 2008, Astronomy \& Astrophysics, 480,
  1

\bibitem[{Noguchi {et~al.}(2021)Noguchi, Hirata, Hirata, Shimaki, Nishikawa,
  Tanaka, Sugiyama, Morota, Sugita, Cho, {et~al.}}]{Noguchi2021}
Noguchi, R., Hirata, N., Hirata, N., {et~al.} 2021, Icarus, 354, 114016

\bibitem[{Roberts {et~al.}(2021)Roberts, Barnouin, Daly, Walsh, Nolan, Daly,
  Michel, Zhang, Perry, Neumann, {et~al.}}]{Roberts2021}
Roberts, J., Barnouin, O., Daly, M., {et~al.} 2021, Planetary and Space
  Science, 204, 105268

\bibitem[{Robinson {et~al.}(2002)Robinson, Thomas, Veverka, Murchie, \&
  Wilcox}]{Robinson2002}
Robinson, M., Thomas, P., Veverka, J., Murchie, S., \& Wilcox, B. 2002,
  Meteoritics \& Planetary Science, 37, 1651

\bibitem[{Rozitis \& Green(2012)}]{Rozitis2012}
Rozitis, B. \& Green, S.~F. 2012, Monthly Notices of the Royal Astronomical
  Society, 423, 367

\bibitem[{Rozitis \& Green(2013)}]{Rozitis2013}
Rozitis, B. \& Green, S.~F. 2013, Monthly Notices of the Royal Astronomical
  Society, 433, 603

\bibitem[{Rubincam(2000)}]{Rubincam00}
Rubincam, D.~P. 2000, Icarus, 148, 2

\bibitem[{Scheeres {et~al.}(2007)Scheeres, Abe, Yoshikawa, Nakamura, Gaskell,
  \& Abell}]{Scheeres2007}
Scheeres, D., Abe, M., Yoshikawa, M., {et~al.} 2007, Icarus, 188, 425

\bibitem[{Scheeres \& Gaskell(2008)}]{Scheeres2008}
Scheeres, D. \& Gaskell, R. 2008, Icarus, 198, 125

\bibitem[{{\v{S}}eve{\v{c}}ek {et~al.}(2015){\v{S}}eve{\v{c}}ek, Bro{\v{z}},
  {\v{C}}apek, \& {\v{D}}urech}]{Sevevcek2015}
{\v{S}}eve{\v{c}}ek, P., Bro{\v{z}}, M., {\v{C}}apek, D., \& {\v{D}}urech, J.
  2015, Monthly Notices of the Royal Astronomical Society, 450, 2104

\bibitem[{{Statler}(2009)}]{Statler2009}
{Statler}, T.~S. 2009, \icarus, 202, 502

\bibitem[{Vincent {et~al.}(2014)Vincent, Schenk, Nathues, Sierks, Hoffmann,
  Gaskell, Marchi, O'Brien, Sykes, Russell, {et~al.}}]{Vincent2014}
Vincent, J.-B., Schenk, P., Nathues, A., {et~al.} 2014, Planetary and Space
  Science, 103, 57

\bibitem[{Vokrouhlicky(1998)}]{Vokrouhlicky1998}
Vokrouhlicky, D. 1998, Astronomy and Astrophysics, 335, 1093

\bibitem[{Vokrouhlick{\`y} {et~al.}(2006)Vokrouhlick{\`y}, Bro{\v{z}}, Bottke,
  Nesvorn{\`y}, \& Morbidelli}]{Vokrouhlicky2006}
Vokrouhlick{\`y}, D., Bro{\v{z}}, M., Bottke, W., Nesvorn{\`y}, D., \&
  Morbidelli, A. 2006, Icarus, 182, 118

\bibitem[{Vokrouhlick{\`y} \& {\v{C}}apek(2002)}]{Vokrouhlicky02}
Vokrouhlick{\`y}, D. \& {\v{C}}apek, D. 2002, Icarus, 159, 449

\bibitem[{Vokrouhlick{\`y} {et~al.}(2004)Vokrouhlick{\`y}, {\v{C}}apek,
  Kaasalainen, \& Ostro}]{Vokrouhlicky2004}
Vokrouhlick{\`y}, D., {\v{C}}apek, D., Kaasalainen, M., \& Ostro, S. 2004,
  Astronomy \& Astrophysics, 414, L21

\bibitem[{Vokrouhlick{\`y} {et~al.}(2000)Vokrouhlick{\`y}, Milani, \&
  Chesley}]{Vokrouhlicky2000}
Vokrouhlick{\`y}, D., Milani, A., \& Chesley, S. 2000, Icarus, 148, 118

\bibitem[{{Vokrouhlick{\'y}} {et~al.}(2003){Vokrouhlick{\'y}}, {Nesvorn{\'y}},
  \& {Bottke}}]{Vokrou2003}
{Vokrouhlick{\'y}}, D., {Nesvorn{\'y}}, D., \& {Bottke}, W.~F. 2003, \nat, 425,
  147

\bibitem[{{Walsh} \& {Jacobson}(2015)}]{WalshJacobson2015}
{Walsh}, K.~J. \& {Jacobson}, S.~A. 2015, in Asteroids IV (University of
  Arizona Press), 375--393

\bibitem[{{Walsh} {et~al.}(2008){Walsh}, {Richardson}, \& {Michel}}]{Walsh2008}
{Walsh}, K.~J., {Richardson}, D.~C., \& {Michel}, P. 2008, \nat, 454, 188

\bibitem[{{Yan} \& {Li}(2019)}]{Yan2019}
{Yan}, X. \& {Li}, J. 2019, Scientia Sinica Physica, Mechanica \& Astronomica,
  49, 084511

\bibitem[{Zhang {et~al.}(2018)Zhang, Richardson, Barnouin, Michel, Schwartz, \&
  Ballouz}]{Zhang18}
Zhang, Y., Richardson, D.~C., Barnouin, O.~S., {et~al.} 2018, The Astrophysical
  Journal, 857, 15

\end{thebibliography}

% \begin{thebibliography}{}

%   \bibitem[Baker(1966)]{baker} Baker, N. 1966,
%       in Stellar Evolution,
%       ed.\ R. F. Stein,\& A. G. W. Cameron
%       (Plenum, New York) 333

%   \bibitem[Balluch(1988)]{balluch} Balluch, M. 1988,
%       A\&A, 200, 58

%   \bibitem[Cox(1980)]{cox} Cox, J. P. 1980,
%       Theory of Stellar Pulsation
%       (Princeton University Press, Princeton) 165

%   \bibitem[Cox(1969)]{cox69} Cox, A. N.,\& Stewart, J. N. 1969,
%       Academia Nauk, Scientific Information 15, 1

%   \bibitem[Mizuno(1980)]{mizuno} Mizuno H. 1980,
%       Prog. Theor. Phys., 64, 544
   
%   \bibitem[Tscharnuter(1987)]{tscharnuter} Tscharnuter W. M. 1987,
%       A\&A, 188, 55
  
%   \bibitem[Terlevich(1992)]{terlevich} Terlevich, R. 1992, in ASP Conf. Ser. 31, 
%       Relationships between Active Galactic Nuclei and Starburst Galaxies, 
%       ed. A. V. Filippenko, 13

%   \bibitem[Yorke(1980a)]{yorke80a} Yorke, H. W. 1980a,
%       A\&A, 86, 286

%   \bibitem[Zheng(1997)]{zheng} Zheng, W., Davidsen, A. F., Tytler, D. \& Kriss, G. A.
%       1997, preprint
% \end{thebibliography}

\appendix
\section{Integration domain for three illumination modes}
\label{appA}
There are three illumination modes for a crater according to different solutions of inequality: {(1) full illumination; (2) two-side illumination; and (3) one-side illumination.} These three illumination modes have different illuminated domains, which  are all equivalent to inequality~(\ref{eq:W}), however. Categorizing them is just for the sake of integration of the total recoil force (see Eq.~(\ref{eq:F})).

To obtain the illuminated area, we need to solve the inequality
\begin{equation}
\label{ieq:cos_phi}
    \cos \phi < \frac{\cos 2 \lambda \cos \theta+ \sin \gamma_0}{\sin 2\lambda \sin \theta}.
\end{equation}
One obvious solution is when the right-hand side of the above inequality is larger than 1,
\begin{equation}
\label{ieq:A2}
    \frac{\cos 2 \lambda \cos \theta+ \sin \gamma_0}{\sin 2\lambda \sin \theta} > 1,
\end{equation}
we have 
\begin{equation}
\label{ieq:cos_2lambda}
    \cos (2\lambda + \theta) > \cos (\frac{\pi}{2} + \gamma_0).
\end{equation}
Noting that $2\lambda + \theta \in (2\lambda, 2\lambda + \pi/2 - \gamma_0)$, $\pi/2 + \gamma_0 \in (0, \pi)$, and the cosine function decreases in this domain, we can easily obtain the condition that inequality (\ref{ieq:A2}) always holds for all $\phi$ and $\theta$, which is given by
\begin{equation}
     \frac{\pi}{2} + \gamma_0 > 2\lambda + \frac{\pi}{2} - \gamma_0 ,
\end{equation}
which gives
\begin{equation}
    \lambda < \gamma_0.
\end{equation}
Therefore, when $\lambda < \gamma_0$, inequality (\ref{eq:W}) holds for all $\theta$ and $\phi$, which means that the crater is illuminated everywhere. This is illumination mode (1). In this case,
\begin{equation}
    \mathcal{W} = \mathcal{Z}.
\end{equation}

When $\lambda > \gamma_0$, the inequality can hold for all $\phi$ when $\theta$ fulfills the requirement according to inequality (\ref{ieq:cos_2lambda})
\begin{equation}
    \theta < \frac{\pi}{2} + \gamma_0 - 2\lambda.
\end{equation}
A hidden condition of the above inequality is $\pi/2 + \gamma_0 - 2\lambda > 0$.
Therefore
\begin{equation}
    \gamma_0 < \lambda < \frac{\pi}{4} + \frac{\gamma_0}{2}.
\end{equation}
This is illumination mode (2). In this mode, when $\theta > \frac{\pi}{2} + \gamma_0 - 2\lambda$, the solution of inequality (\ref{ieq:cos_phi}) is $\phi_1 < \phi < \phi_2$. Therefore, illumination mode (2) can be described as 
\begin{equation}
\begin{aligned}
    \mathcal{W} &= \{ (x,y,z) \in \mathcal{Z} \lvert  \theta \in (0,\frac{\pi}{2}-2\lambda+\gamma_0), \phi \in (0,2 \pi) \\ 
    & {\rm or} \,\,\theta \in (\frac{\pi}{2}-2\lambda+\gamma_0 , \frac{\pi}{2} - \gamma_0), \phi \in (\phi_1 ,\phi_2) \}.
\end{aligned}
\end{equation}

For $\lambda > \pi/4 + \gamma_0/2$, the solution of inequality (\ref{ieq:cos_phi}) is $\phi_1 < \phi < \phi_2$, but we have to ensure that the right-hand side is larger than -1,
\begin{equation}
    \frac{\cos 2 \lambda \cos \theta+ \sin \gamma_0}{\sin 2\lambda \sin \theta} > -1,
\end{equation}
which yields
\begin{equation}
    \theta < \frac{\pi}{2} + \gamma_0 - 2\lambda.
\end{equation}
Therefore, illumination mode (3) is equivalent to
\begin{equation}
    \mathcal{W} = \{ (x,y,z) \in \mathcal{Z} \lvert  \theta \in (2\lambda - \gamma_0 - \pi/2, \pi/2 - \gamma_0), \phi \in (\phi_1 ,\phi_2)\}.
\end{equation}

%  Therefore, the illuminated region is 
% % \begin{equation}
% %     \left\{
% %     \begin{aligned}
    
% %     \end{aligned}
% %     \right.
% % \end{equation}

% When $\lambda > \pi/4 +\gamma_0/2$，Inequality (\ref{ieq:A1}) is never valid. But we need to make sure 
% \begin{equation}
%     \frac{\cos 2 \lambda \cos \theta+ \sin \gamma_0}{\sin 2\lambda \sin \theta} > -1,
% \end{equation}
% which gives
% \begin{equation}
%     \cos (2\lambda - \theta) > \cos (\frac{\pi}{2} + \gamma_0).
% \end{equation}
% Therefore
% \begin{equation}
%     \theta > 2 \lambda - \frac{\pi}{2} - \gamma_0.
% \end{equation}

\section{$\delta$ of $Z$-axis symmetric asteroids}
\label{appB}
The variable $\delta$ is the longitude difference between the position vector $\vec r_0$ and the normal vector $\vec n_0$. We show below that when the shape of the asteroid is $Z$-axis symmetric, $\delta = 0$ is valid everywhere on the surface of the asteroid.

For an arbitrary $Z$-axis symmetric asteroid, the unit position vector of a point on the surface can be expressed by two independent variables $\zeta$ and $k$ as
\begin{equation}
\label{eq:r_0_1}
    \vec r_0 = (k \cos \zeta, k \sin \zeta, p(k)),
\end{equation}
where $p(k)$ is a function of $k$, depending on the specific shape of the asteroid. If the asteroid is an unit sphere, for example, we have $p(k) = \pm \sqrt{1 - k^2}$. The unit normal vector is 
\begin{equation}
\label{eq:n_0_1}
\begin{aligned}
    \vec n_0 & =  \frac{\partial \vec r_0}{\partial \zeta} \times    \frac{\partial \vec r_0}{\partial k}    \\
    & = - \frac{ {\rm d}p /{\rm d} k}{\sqrt {1 + ({\rm d}p /{\rm d} k)^2}} (\cos \zeta, \sin \zeta, -\frac{1}{{\rm d}p /{\rm d} k} ).   
\end{aligned}
\end{equation}
By comparing Equation~(\ref{eq:r_0_1}) and (\ref{eq:n_0_1}), we see that $\vec r_0$ and $\vec n_0$ share the same azimuth angle $\zeta$. Therefore, for any point on a $Z$-axis symmetric asteroid, the difference between the azimuth angles of the position vector and the normal vector is zero, which means $\delta = 0$ in Equation (\ref{eq:n_0}).

% For the first case, the whole crater can be illuminated and there is no shadow. This requires $\lambda < \gamma_0$, and it can be seen from Inequity (\ref{eq:W}). Since $\theta < \pi/2 - \gamma_0$, the right side in Inequity (\ref{eq:W}) is
% \begin{equation}
%     \frac{\cos 2 \lambda \cos \theta+ \sin \gamma_0}{\sin 2\lambda \sin \theta} > \frac{\cos 2 \lambda \sin \gamma_0 + \sin \gamma_0}{\sin 2\lambda \cos \gamma_0} > 1,
% \end{equation}
% which makes Inequity (\ref{eq:W}) always holds, implying that everywhere in the crater is illuminated. Therefore, $\theta \in (0, \pi/2 - \gamma_0)$, and $\phi \in (0, 2 \pi)$.

% For the second case, two sides of the crater are illuminated, implying that in the bottom of the crater, there is an illuminated region where $\phi \in (0,\pi)$. An equivalent mathematical description is that $\exists \theta \in (0, \pi/2-\gamma_0)$ such that Inequality \ref{eq:W} holds for $\phi \in (0, 2\pi)$. This is also equivalent to the argument that
% \begin{equation}
%     \frac{\cos 2 \lambda \cos \theta+ \sin \gamma_0}{\sin 2\lambda \sin \theta} > 1
% \end{equation}
% always holds, which implies
% \begin{equation}
%     \left\{
%     \begin{aligned}
%     \cos 2 \lambda \cos \theta+ \sin \gamma_0 > 
%     \end{aligned}
    
%     \right.
% \end{equation}

\end{document}